%% file: main.tex
\documentclass[a4paper,twocolumn,10pt]{quantumarticle}
\pdfoutput=1

\usepackage[utf8]{inputenc}
\usepackage[english]{babel}
\usepackage[T1]{fontenc}
\usepackage{amsmath}
\usepackage{hyperref}
\usepackage{tikz}

\usepackage{cite}
\usepackage{amsmath,amssymb,amsfonts}
\usepackage{graphicx}
\usepackage{textcomp}
\usepackage{xcolor}
\usepackage{mathtools}
\usepackage[ruled,noend]{algorithm2e}

\usepackage{microtype}
\usepackage{float}
\usepackage{adjustbox}
\usepackage{booktabs,makecell,tabularx}
\usepackage{url}
\usepackage{booktabs}
\usepackage{caption}
\usepackage{subcaption}
\usepackage[braket, qm]{qcircuit}

\usepackage{tikz}
\usepackage{preview}
\newcounter{row}
\newcounter{col}

\definecolor{refblue}{RGB}{102, 102, 153}
\definecolor{stringgreen}{RGB}{80, 107, 65}
\definecolor{keywordblue}{RGB}{64, 89, 245}
\definecolor{commentbrown}{RGB}{59, 35, 0}

\usepackage{listings}
\lstdefinestyle{lststyle}{
  commentstyle=\color{commentbrown},
  keywordstyle=\color{keywordblue},
  numberstyle=\tiny\color{gray},
  stringstyle=\color{stringgreen},
  basicstyle=\ttfamily\footnotesize,
  breakatwhitespace=false,
  breaklines=true,
  numbers=none,
  captionpos=b,
  frame=lines,
  keepspaces=true,
  numbers=left,
  numbersep=5pt,
  showspaces=false,
  showstringspaces=false,
  showtabs=false,
  tabsize=1,
  columns=fullflexible 
}
\newcommand\setrow[4]{
  \setcounter{col}{1}
  \foreach \n in {#1, #2, #3, #4} {
    \edef\x{\value{col} - 0.5}
    \edef\y{4.5 - \value{row}}
    \node[anchor=center,font=\Large] at (\x, \y) {\n};
    \stepcounter{col}
  }
  \stepcounter{row}
}

\newcolumntype{C}[1]{>{\centering\arraybackslash}p{#1}}
\newcolumntype{L}{>{\raggedright\arraybackslash}X}
\usepackage{siunitx}
\usepackage[utf8]{inputenc}
\renewcommand{\thetable}{\arabic{table}}
\usepackage{etoolbox}
\usepackage{xparse}
\makeatletter

\newcommand{\I}{\mathbb I}
\newcommand{\CZ}{\mathrm{CZ}}
\newcommand{\CX}{\mathrm{CX}}
\newcommand{\MCZ}{\mathrm{MCZ}}
\newcommand{\XXYY}{\mathrm{XX+YY}}

\newcommand{\nnot}{\texttt{not}\,}
\newcommand{\accept}{\texttt{accept}}
\newcommand{\reject}{\texttt{reject}}
\newcommand{\oddity}{\texttt{oddity}}

\newcommand{\varaccept}{\texttt{accept\_qbl}}
\newcommand{\varreject}{\texttt{reject\_qbl}}
\newcommand{\varoddity}{\texttt{oddity\_qbl}}
\newcommand{\vareven}{{even}}
\newcommand{\varh}{\texttt{h}}
\newcommand{\varbranchqa}{\texttt{branch\_qa}}
\newcommand{\varbranchqv}{\texttt{branch\_qv}}

\newcommand{\varmaxdepth}{{max\_depth}}
\newcommand{\vardegree}{{deg}}

\newcommand{\varctrl}{\texttt{ctrl}}
\newcommand{\vartempqv}{\texttt{temp\_qv}}

\SetKwFor{Control}{with control}{do}{end}
\SetKwBlock{Invert}{with invert}{end}
\SetKw{KwBy}{by}
\SetKwBlock{KwMCZ}{MCZ}{end}
\SetKw{KwCX}{CX}
\SetKw{KwXXYY}{XX+YY}
\SetKw{KwH}{H}
\SetKw{KwSwap}{SWAP}
\renewcommand{\tcp}[1]{\textit{/*#1}\break}
\SetKw{KwContinue}{continue} 

\begin{document}

\title{Quantum Backtracking in Qrisp Applied to Sudoku Problems}

\author{Raphael Seidel}
\email{raphael.seidel@fokus.fraunhofer.de}
\affiliation{Fraunhofer Institute for Open Communication Systems, Berlin, Germany}

\author{René Zander}
\email{rene.zander@fokus.fraunhofer.de}
\affiliation{Fraunhofer Institute for Open Communication Systems, Berlin, Germany}

\author{Matic Petri\v{c}}
\email{matic.petric@fokus.fraunhofer.de}
\affiliation{Fraunhofer Institute for Open Communication Systems, Berlin, Germany}

\author{Niklas Steinmann}
\email{niklas.steinmann@fokus.fraunhofer.de}
\affiliation{Fraunhofer Institute for Open Communication Systems, Berlin, Germany}

\author{David Q. Liu}
\email{dqliu@purdue.edu}
\affiliation{Purdue University, West Lafayette, IN, USA}

\author{Nikolay Tcholtchev}
\email{nikolay.tcholtchev@fokus.fraunhofer.de}
\affiliation{Fraunhofer Institute for Open Communication Systems, Berlin, Germany}

\author{Manfred Hauswirth}
\email{manfred.hauswirth@fokus.fraunhofer.de}
\affiliation{Fraunhofer Institute for Open Communication Systems, Berlin, Germany}
\affiliation{Technische Universität Berlin, Berlin, Germany}


\maketitle

\begin{abstract}
  The quantum backtracking algorithm proposed by Ashley Montanaro raised considerable interest, as it provides a quantum speed-up for a large class of classical optimization algorithms. It does not suffer from Barren-Plateaus and transfers well into the fault-tolerant era, as it requires only a limited number of arbitrary angle gates. Despite its potential, the algorithm has seen limited implementation efforts, presumably due to its abstract formulation. In this work, we provide a detailed instruction on implementing the quantum step operator for arbitrary backtracking instances. For a single controlled diffuser of a binary backtracking tree with depth $n$, our implementation requires only $6n+14$ CX gates. We detail the process of constructing accept and reject oracles for Sudoku problems using our interface to quantum backtracking. The presented code is written using Qrisp, a high-level quantum programming language, making it executable on most current physical backends and simulators. Subsequently, we perform several simulator based experiments and demonstrate solving 4x4 Sudoku instances with up to 9 empty fields. This is, to the best of our knowledge, the first instance of a compilable implementation of this generality, marking a significant and exciting step forward in quantum software engineering.
\end{abstract}

\input{sections/introduction}

\input{sections/QuantumBacktracking}

\section{Quantum Backtracking Implementation}
\label{sec:backtracking_implementation}
\input{sections/backtracking_implementation}
\input{sections/psi_prep}

\input{sections/oracle_implementation}

\input{sections/Experiments}

\input{sections/Conclusion}

\section{Acknowledgment}
\input{sections/Acknowledgment}

\section*{Data availability}
All data generated or analysed during this study are included in this published article.

\section*{Code availability}
\textit{Qrisp} is an open-source python framework for high-level programming of quantum computers.
The source code is available in \href{https://github.com/eclipse-qrisp/Qrisp}{https://github.com/eclipse-qrisp/Qrisp}.
A tutorial explaining the implementation of the Sudoku solver can be found on the webpage \href{https://www.qrisp.eu/general/tutorial/Sudoku.html}{https://www.qrisp.eu/general/tutorial/Sudoku.html}.

\bibliographystyle{quantum}
\bibliography{sources}

\appendix
\input{sections/appendix}

\end{document}

%% file: sections/introduction.tex
\section{Introduction}
The field of quantum optimization \cite{Abbas_2023} had to suffer several significant blows in recent years. One by one, several promising algorithms became the target of serious doubts \cite{Barak_2015, Tang_2019, Stoudenmire_2023, Cerezo_2023}. Despite these challenges, the ubiquitous prevalence of optimization problems in industrial settings suggests that even minor speed-ups could have a comparable impact to exponential algorithms, which are usually limited in their range of application.
The quantum backtracking algorithm for solving constrained satisfaction problems treated here \cite{Montanaro_2016} offers a near-quadratic speed-up over a classical backtracking algorithm. Moreover, it was generalized to similarly accelerate classical branch-and-bound algorithms for constrained optimization problems \cite{Montanaro_2020}.


In this work, we present a systematic and detailed implementation of Montanaro's quantum backtracking algorithm \cite{Montanaro_2016}, subsequently applying it to Sudoku problems (Figure \ref{fig:sudoku}) to demonstrate its practicality. The source code is available in \href{https://github.com/eclipse-qrisp/Qrisp}{https://github.com/eclipse-qrisp/Qrisp}.

\PreviewEnvironment{tikzpicture}
\begin{figure}[htbp]
\centering
\begin{tikzpicture}[scale=1.0]

  \begin{scope}
    \draw (0, 0) grid (4, 4);
    \draw[very thick, scale=2] (0, 0) grid (2, 2);

    \setcounter{row}{1}
    \setrow {1}{}{3}{}
    \setrow {3}{}{1}{}

    \setrow {}{1}{}{3}
    \setrow {4}{}{}{}

  \end{scope}

\end{tikzpicture}
\caption{Unsolved 4x4 Sudoku problem: A solution to this Sudoku with 9 empty fields is found on a simulator utilizing the \textit{Qrisp} implementation of Montanaro's algorithm with 91 qubits and a circuit depth of 3968.
}
\label{fig:sudoku}
\end{figure}

\subsection{Qrisp}
In the original publication of the algorithm, the author refrains from providing any specific details regarding its implementation. As a matter of fact, there is little to no literature on how such an implementation, at least in terms of quantum circuits, could look like \cite{Dalzell_2023, Martiel_2020}.
This is likely due to the rather complex requirements of the quantum step operator, which pose significant challenges when it comes to software engineering.

To tackle such challenges, the high-level programming framework \textit{Qrisp} has been developed over the past several years. One of the defining standout features of \textit{Qrisp} is its ability of eliminating the need for direct manipulation of quantum circuits during software development. Despite this, \textit{Qrisp} remains fully compilable, while also offering the convenience of importing and exporting quantum circuits to other popular representations \cite{qiskit,tket,pennylane}.

Why is it beneficial to use a high-level programming language like \textit{Qrisp} for  the construction of complex quantum algorithms? Answering this can be challenging with the lack of metrics being able to effectively capture the advantages of a well-structured codebase. Nevertheless, classical high-level programming languages are flourishing more than ever with their seemingly undeniable advantages. We identify the following key points:

\begin{enumerate}
\item \textbf{Variables and Functions instead of Qubits and Gates}. Using variables instead of qubits helps structuring the code immensely while also setting the basis for a plethora of established programming concepts adapted to the field of quantum computing. Apart from the advantages in the codebase, using these concepts allows users to start THINKING in a quantum way that resembles a programmer, instead of someone chaining together unitary matrices. This is because the code written in \texttt{QuantumVariable} notation tends to be significantly shorter and is able to capture the programmer's intent more clearly. Cumbersome tasks like uncomputing garbage qubits are eliminated due to a wide range of automations, enabling the programmer to focus on the essential elements of their work.
\item \textbf{Modularity}. It cannot be overstated that modularity is an inevitable feature for any future-proof codebase. What do we mean by this? Modularity is a property that groups the code into separate organizational units, which interact less. It therefore allows a team of programmers to work on different parts of the code without having to frequently reach agreements. As an immediate consequence, there can be separate domain experts working together. Modularity also facilitates a simple replacement of modules with potential alternatives. Finally, it allows testing and/or benchmarking modules in isolated environments, which is crucial for bug and/or bottleneck identification. 
Among the established Python SDKs \cite{qiskit,cirq,tket,Efthymiou_2021,pennylane}, \textit{Qrisp} is the only framework enabling true modularity. This is because quantum memory (i.e. qubits) has to be managed \textit{manually} in circuit based representations. Exchanging a module to an alternative one, which requires a different amount of temporary ancilla qubits, implies adjusting the whole codebase to properly recycle the newly allocated resources. In \textit{Qrisp}, this task is automated due to the embedding in the \texttt{QuantumVariable} system. As an example, the \textit{Qrisp} implementation of Shor's algorithm can be called effortlessly using different kinds of adders, ranging from slow adders requiring no additional ancillae \cite{Cuccaro_2004} to fast adders with a higher demand \cite{Gidney_2018, Wang_2023}.

\item \textbf{Platform independence}. The advantage of running code on a variety of backends is rather self-explanatory. Instead, we want to highlight why many frameworks, despite their claims of possessing such features, indeed lack them. Platform independence doesn't mean that it is possible to convert the compilation result to many different circuit representations. This is because, for most backends, there are much better device-specific implementations of elementary functions\cite{Seidel_2022}. \textit{Qrisp} supports a variety of circuit conversion functions, but the true platform independence comes from its architecture: Because of its dynamic qubit management, the implementation of virtually any elementary function can be replaced such that the same code compiles to an entirely different circuit.\\
\end{enumerate}

\subsection{Related work}
\subsubsection{Quantum backtracking}
In the past years a few extensions have been proposed. Crucially, in \cite{Ambainis_2017} an algorithm for tree size estimation was given. This is relevant, because the original publication alone made no explicit statements about how the precision should be chosen. With the tree size determination algorithm at hand, a more sophisticated method of precision estimation is available. In \cite{Montanaro_2020} Montanaro generalized his algorithm to branch and bound methods, enabling an even wider field of applications. The following papers applied the algorithm:

\begin{itemize}
    \item In \cite{Aono_2018} the algorithm is applied to quantum lattice enumeration.
    \item \cite{Martiel_2020} applies the algorithm for general CSP problems.
    \item In \cite{Montanaro_2017} an application to solve the traveling salesman problem is described.
    \item Exact satisfiability is tackled in \cite{Mandra_2016}.
\end{itemize}

The speed-up of a physical implementation of the algorithm, including hardware parameters like timing, error-rates etc., has been studied in \cite{Campbell_2019}.

\subsubsection{High level programming}

Establishing a practically useful environment for quantum software engineering is an ongoing effort. The outstanding problem here is to find abstractions that simplify the programming workflow and enable systematic software development, while not preventing programmers from expressing arbitrary quantum operations. Over the past years several high-level languages have been proposed. Many of them fall however either into the category of not being able to compile to the circuit level (in a finite amount of time) \cite{Bichsel_2020, Voichick_2023, Steiger_2018} or still being quite close to the circuit level \cite{Svore_2018,Green_2013, Guo_2023, qiskit, tket, pennylane, cirq}. Two representatives that live on less extreme ends of the spectrum can be found in \cite{Yuan_2022}, \cite{Varga_2024}.

The next burden of quantum language design is the cost of abstraction: As quantum computing resources are very limited for both physical backends and simulators, the compiled circuits need to be able to make use of every available shortcut. Usually this implies that abstractions have to be broken, demotivating the use of said abstractions in the first place. This is true only until a certain level of complexity is reached. Beyond that point, manual optimization might yield more efficient circuits, but these become practically unachievable due to the scale of the problem. This is the point where the advantages of design automation kick in. A strong example for this phenomenon has been given in \cite{Paradis_2021}, which describes an algorithm that achieves the automatic uncomputation of a wide class of quantum circuits. Instead of simply uncomputing, the algorithm utilizes "garbage phases" canceling each other, thus relaxing the requirements to the compiler, resulting in more efficient gates (for more details of this phenomenon see section \ref{sec:pt_compilation}). The proposed uncomputation algorithm has been integrated in \textit{Qrisp} and proved to be useful for many functions of this work. While the user has to manually specify in some situations that a phase tolerant replacement is a valid transformation, the process is automated in most other cases.

\subsection{Paper organization}
This paper is organized as follows:
A brief introduction to classical backtracking for solving constrained satisfaction problems, together with a description of Montanaro's quantum backtracking algorithm, is provided in Section \ref{sec:quantum_backtracking}.
In Section \ref{sec:backtracking_implementation}, we explain the implementation of Montanaro's algorithm in \textit{Qrisp}. Here we provide a detailed description of the $\texttt{QuantumBacktrackingTree}$ class, and the belonging methods for implementing the quantum walk on the backtracking tree.
The implementation of the oracle functions tailored to the Sudoku problem is explained in Section \ref{sec:oracle_implementation}.
Experiments and benchmarking are discussed in Section \ref{sec:experiments}.
The Appendices cover various aspects of the optimization of controlled algorithmic primitives and other compilation specifics.

%% file: sections/QuantumBacktracking.tex
\section{Quantum Backtracking}
\label{sec:quantum_backtracking}

In this section, we provide a brief introduction to classical backtracking and Montanaro's quantum backtracking algorithm.

\subsection{Classical backtracking}
A constraint satisfaction problem (CSP) can be described as follows: given a predicate $\accept\colon [d]^n\rightarrow\{\texttt{True},\texttt{False}\}$ where $[d]=\{0,\dotsc,d-1\}$, find an assignment $x$ to the $n$ variables such that $\accept(x)$ is \texttt{True}, or output ``no solution" if no such $x$ exists.
This general class of problems encompasses, e.g., the Boolean satisfiability problem (SAT) and the graph colouring problem. 

Backtracking is a well-known classical method for solving CSPs and thereby taking advantage of their structure. This approach can be utilized whenever we have the ability to recognize whether partial solutions to a problem can be extended to full solutions. For this, assume predicate functions $\accept$ and $\reject$ that can be evaluated on a partial assignment $x\in\mathcal D$ where $\mathcal D=([d]\cup\{*\}])^n$. Here, the $*$'s represent unassigned values. An assignment $x$ is \textit{complete} if it contains no $*$'s. The predicate $\accept(x)$ returns \texttt{True} if $x$ is a solution to the CSP (we say $x$ is marked), and \texttt{False} otherwise. 
The predicate $\reject(x)$ returns $\texttt{True}$ if it is clear that $x$ cannot be extended to a solution, and $\texttt{False}$ otherwise. 
Additionally, we assume access to a heuristic $h\colon\mathcal D\rightarrow\{1,\dotsc,n\}$ that determines which unassigned value of $x=(x_1,\dotsc,x_n)$ to assign in the next step. Finally, classical backtracking can be implemented as described in Algorithm \ref{alg:classical_bt}. 
We can construct Algorithm \ref{alg:classical_bt} as exploring a tree $\mathcal T$ whose internal nodes are partial assignments, and whose leaves are solutions to $P$, i.e., $\accept(x)=\texttt{True}$ ($x$ is marked), or cannot be extended to a solution, i.e., $\reject(x)=\texttt{True}$ or $x$ is complete with $\accept(x)=\texttt{False}$.

\begin{algorithm}[h]
\caption{Classical backtracking}\label{alg:classical_bt}

\KwIn{A partial assignment $x \in \mathcal{D}$, and predicate functions $\accept, \reject : \mathcal{D} \rightarrow \{\texttt{True}, \texttt{False}\}$, and a heuristic $h : \mathcal{D} \rightarrow \{1, \dotsc, n\}$.}
\KwOut{Solutions to the CSP.}

\If{$\accept(x) == \texttt{True}$}{
    \KwRet $x$\;
}
\If{$\reject(x) == \texttt{True}$ \textbf{or} $x$ is a complete assignment}{
    \KwRet\;
}
$i \gets h(x)$\;
\For{$w \in [d]$}{
    $y \gets x$ with $i$th entry replaced by $w$\;
    Algorithm \ref{alg:classical_bt}$(y, \accept, \reject, h)$\;
}
\end{algorithm}

\subsection{Montanaros's algorithm}
In this part, we describe Montanaro's backtracking algorithm \cite{Montanaro_2016} for detecting a marked node in a tree. The algorithm is based on a quantum walk starting at the root of a backtracking tree.

Consider a rooted tree $\mathcal T$ with $T$ nodes $r,1,\dotsc T-1$, with $r$ being the root of $\mathcal T$. The depth of the tree, i.e., the maximal distance from the root to any leaf, is denoted by $n$. For $x\in\mathcal T$, we denote the subtree with root $x$ by $\mathcal T_x$.
Let $A$ be the set of nodes with an even distance from the root (including the root itself), and let $B$ be the set of nodes with and odd distance from the root. We write $x\rightarrow y$ to indicate that $y$ is a child of $x$ in the tree. 
The quantum walk operates on the Hilbert space $\mathcal H$ spanned by $\{\ket{r}\}\cup\{\ket{x}\mid x\in 1,\dotsc T-1\}$ and starts in the state $\ket{r}$. It is based on the diffusion operators $D_x$ that act on the 
subspace $\mathcal H_x$ spanned by $\{\ket{x}\}\cup\{\ket{y}\mid x\rightarrow y\}$. The diffusion operators are defined as follows: if $x$ is marked, then $D_x=\I$. If $x$ is not marked, then $D_x=\I-2\ket{\psi_x}\bra{\psi_x}$ where 
\begin{equation}
\ket{\psi_x}\propto\ket{x}+c\sum\limits_{y,\,x\rightarrow y}\ket{y}
\end{equation}
with $c=\sqrt{n}$ if $x=r$ and $c=1$ otherwise. Note that $D_x=-\I$ if $x$ is a leaf that is not marked (i.e., if $x$ cannot be extended to a solution).
With this, a step of the quantum walk consists of applying the operator $R_BR_A$ where 
\begin{equation}
R_A=\bigoplus_{x\in A}D_x\quad\text{and}\quad R_B=\ket{r}\bra{r}+\bigoplus_{x\in B}D_x.
\end{equation}
Then the presence of a marked node in the tree $\mathcal T$ can be detected as described in Algorithm \ref{alg:detect_solution}, and with this a marked node can be found efficiently via binary search \cite{Montanaro_2016}. 

\begin{algorithm}[h]
\caption{Detecting a marked node}\label{alg:detect_solution}

\KwIn{Operators $R_A, R_B$, a failure probability $\delta$, upper bounds on the depth $n$ and the number of nodes $T$, and universal constants $\beta, \gamma>0$ to be determined.}
\KwOut{``marked node exists" or ``no marked node"}

$accept\_number \gets 0$\;
\For{$i\gets 1$ \KwTo $K=\lceil\gamma\log(1/\delta)\rceil$}{
    apply phase estimation to $R_BR_A$ with precision $\beta/\sqrt{Tn}$\;
    \If{the eigenvalue is $1$}{
        $accept\_number \gets  accept\_number+1$\;
    }
}
\If{$accept\_number\geq3K/8$}{
    \Return ``marked node exists"\;
}
\Else{
    \Return ``no marked node"\;
}
\end{algorithm}

%% file: sections/backtracking_implementation.tex
In this section, we describe the implementation of Montanaro's backtracking algorithm in \textit{Qrisp}.
We use the detection of a marked vertex by applying quantum phase estimation to the operator $R_BR_A$ as described in Algorithm \ref{alg:detect_solution}. As part of the \texttt{QuantumBacktrackingTree} class, the operator $R_BR_A$ is implemented by the \texttt{quantum\_step} function consisting of two steps of the \texttt{qstep\_diffuser} corresponding to $R_A$ and $R_B$, respectively. Phase estimation with a specified precision is then applied by the \texttt{estimate\_phase} function. The \texttt{find\_solution} method starts by applying the \texttt{estimate\_phase} function to the entire tree (initialized in the state $\ket{r}$), and then, based on the measurement results, recursively applies the function to subtrees in order to find a solution.
In the following, we provide a detailed description of the \texttt{QuantumBacktrackingTree} class, and the implementation of the \texttt{qstep\_diffuser} and the belonging auxiliary methods.

\subsection{The \texttt{QuantumBacktrackingTree} class}

This class describes the central data structure for the implementation of quantum backtracking algorithms based on the use of a discrete time quantum walk.
The class includes, e.g, the $\texttt{qstep\_diffuser}$ function which implements the quantum walk on the tree.

An instance of the $\texttt{QuantumBacktrackingTree}$ class requires the specification of four parameters during initialization:
\begin{itemize}
    \item An integer $\varmaxdepth$ specifying the depth of the backtracking tree.
    \item A \texttt{QuantumVariable} $\varbranchqv$ representing the possible branches of each node.
    \item An $\accept$ function that returns \texttt{True} if called on a node that is a solution, and \texttt{False} otherwise.
    \item A $\reject$ function that returns \texttt{True} if called on a node $x$ representing a partial solution for which it is clear that $x$ cannot be extended to a solution, and \texttt{False} otherwise.
\end{itemize}
The \texttt{accept} and \texttt{reject} functions allow the algorithm to quickly discard large parts of the potential solution space by using the $\texttt{reject}$ function to cancel the recursion. 
The predicate functions \texttt{accept} and \texttt{reject} must meet further conditions
for the algorithm to function correctly:
\begin{enumerate}
    \item Both functions have to return a \texttt{QuantumBool}
    \item Both functions must not change the state of the tree.
    \item Both functions must delete (uncompute) all temporarily created \texttt{QuantumVariables}.
    \item The \texttt{accept} and \texttt{reject} functions must never return \texttt{True} on the same node.
\end{enumerate}

An instance of the \texttt{QuantumBacktrackingTree} class represents tree of depth $\varmaxdepth$ such that every node either has either 0 children (if it is rejected) or $\vardegree=2^m$ children otherwise, where $m$ denotes the size of the \texttt{QuantumVariable branch\_qv}. Notably, this differs from the backtracking tree $\mathcal T$ as described in Section \ref{sec:quantum_backtracking} where nodes can have an arbitrary number $\leq 2^m$ of children depending on how many are rejected. At the cost of increasing the search space by including also the rejected nodes, this change enables a much more efficient \texttt{quantum\_step} procedure, as elaborated below.

Additionally, the class exhibits the following attributes that allow for the encoding of (superpositions of) node states $\ket{x}$:
\begin{itemize}
    \item A \texttt{QuantumVariable} $\varh$ representing the height of the node $x$ (e.g., a leaf has height 0 and the root has height $\varmaxdepth$) as a one-hot encoded integer. That is, $\varh$ consists of a register of $\varmaxdepth+1$ qubits, and 
    heights are encoded by the states where exactly one qubit is $1$. 
    \item A \texttt{QuantumArray} $\varbranchqa$ representing the path from the root to the node $x$. The array has size $\varmaxdepth$ and the entries are of the type $\varbranchqv$. 
\end{itemize}
A node state $\ket{x}$ is encoded as the combination of $\ket{\varh}$ and $\ket{\varbranchqa}$: 
\begin{equation}
\ket{x} = \ket{\varbranchqa}\ket{\varh}. 
\end{equation}
An example of encoding the node with path [0,1] in a binary tree of depth 4 is
\begin{align*}
\ket{\varbranchqa} &= \ket{[0,0,1,0]}\\
\ket{\varh} &= \ket{2} = \ket{00100}\\
\ket{x} &= \ket{\varbranchqa}\ket{\varh}
\end{align*}
Note that the state of the array $\varbranchqa$ is the reversed path. States that have non-zero values at entries indexed smaller than $\varh$ are considered as non-algorithmic, e.g., $\ket{\varh}=\ket{2}$ and $\ket{\varbranchqa}=\ket{[0,1,1,1]}$.

It is important to note that our specification of a node by its height differs from the encoding proposed in Montanaro's paper, which uses the distance from the root $\ell(x)$. This decision was made to ensure that the value of the parameter does not change when subtrees are considered.

We want to explain this data structure using simple example $\texttt{accept}$ and $\texttt{reject}$ functions, and a visualization of the resulting tree. The $\texttt{accept}$ function below simply marks the node with the path [1,1,1] starting from the root in a tree of depth 3: 
\begin{lstlisting}[language = Python, numbers = none]
@auto_uncompute
def accept(tree):
    return (tree.branch_qa[0] == 1) & 
           (tree.branch_qa[1] == 1) &
           (tree.branch_qa[2] == 1)
\end{lstlisting}
The $\texttt{reject}$ function discards the node with path [0] starting from the root in a tree of depth 3:
\begin{lstlisting}[language = Python, numbers = none]
@auto_uncompute
def reject(tree):
    return (tree.h==2) & 
           (tree.branch_qa[2] == 0)
\end{lstlisting}

Given the \texttt{accept} and \texttt{reject} functions described above, the code for setting up the \texttt{QuantumBacktrackingTree} could be given as: 

\begin{lstlisting}[language = Python, numbers = none]
from qrisp import *
from qrisp.quantum_backtracking
     import QuantumBacktrackingTree as QBT

tree = QBT(max_depth = 3,
           branch_qv = QuantumFloat(1),
           accept = accept,
           reject = reject)
           
tree.init_node([])
tree.visualize_statevector()
\end{lstlisting}

\begin{figure}[htbp]
\centering
\includegraphics[scale=0.4,trim={0cm 0cm 0cm 0cm}]{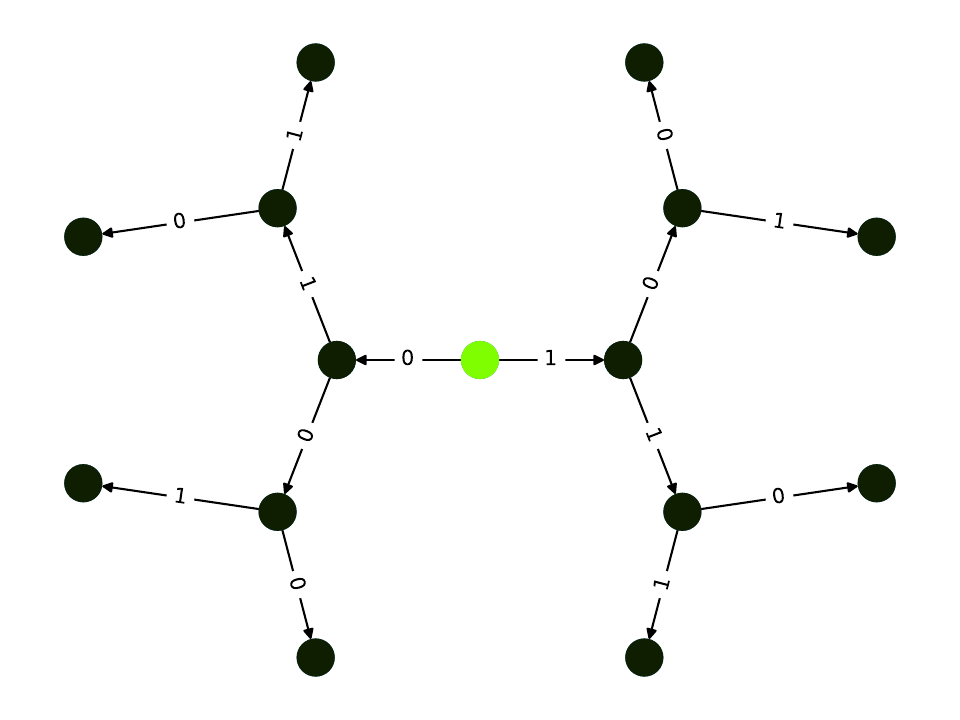}
\caption{Visualisation of a depth 3 \texttt{QuantumBacktrackingTree} initialized in the state $\ket{r}$ (green).}
\label{fig:TreeDepthThree}
\end{figure}

The \texttt{visualize\_statevector} method facilitates displaying the resulting tree of depth 3 initialized in the state $\ket{r}$ as shown in Fig.~\ref{fig:TreeDepthThree}. 

\begin{figure*}[t]
    \begin{subfigure}[]{0.5\textwidth}
        \centering
        \includegraphics[scale=0.4,trim={0.5cm 0cm 0cm 0cm}]{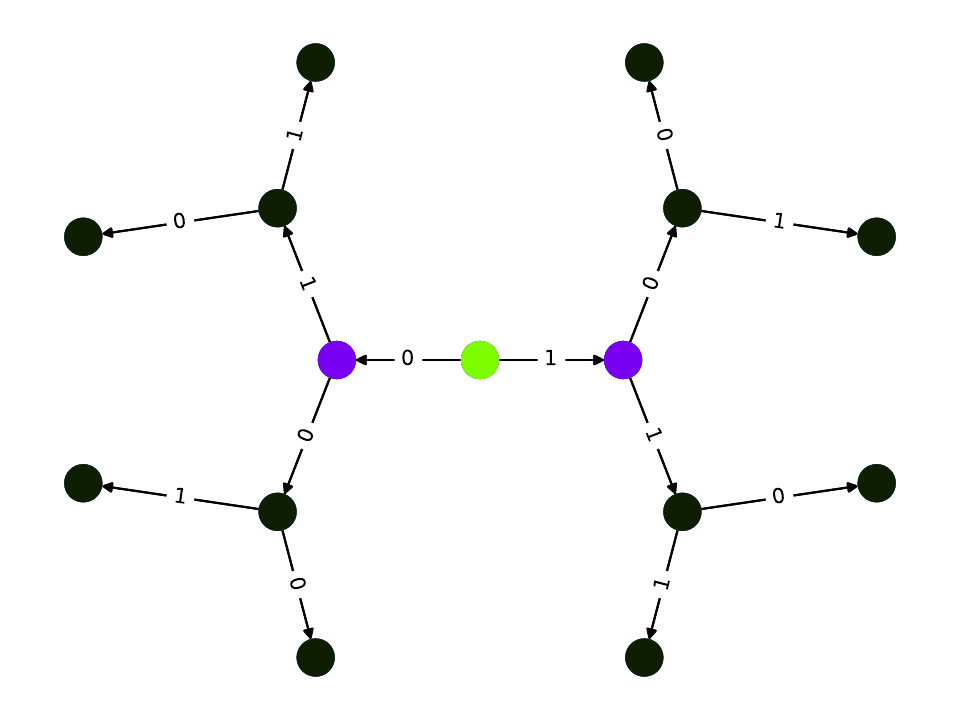}
        \subcaption{$R_A\ket{r}$}
        \label{fig:TreeDepthThree_1A}
    \end{subfigure}
    \begin{subfigure}[]{0.5\textwidth}
        \centering
        \includegraphics[scale=0.4,trim={0.5cm 0cm 0cm 0cm}]{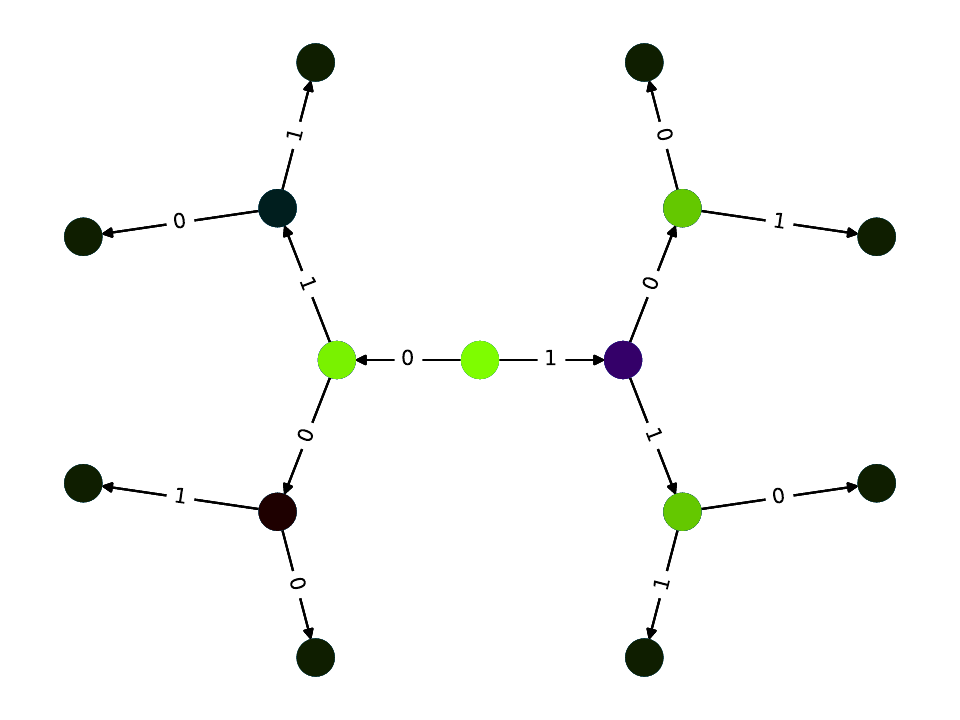}
        \subcaption{$R_BR_A\ket{r}$}
        \label{fig:TreeDepthThree_1B}
    \end{subfigure}

    \begin{subfigure}[]{0.5\textwidth}
        \centering
        \includegraphics[scale=0.4,trim={0.5cm 0cm 0cm 0cm}]{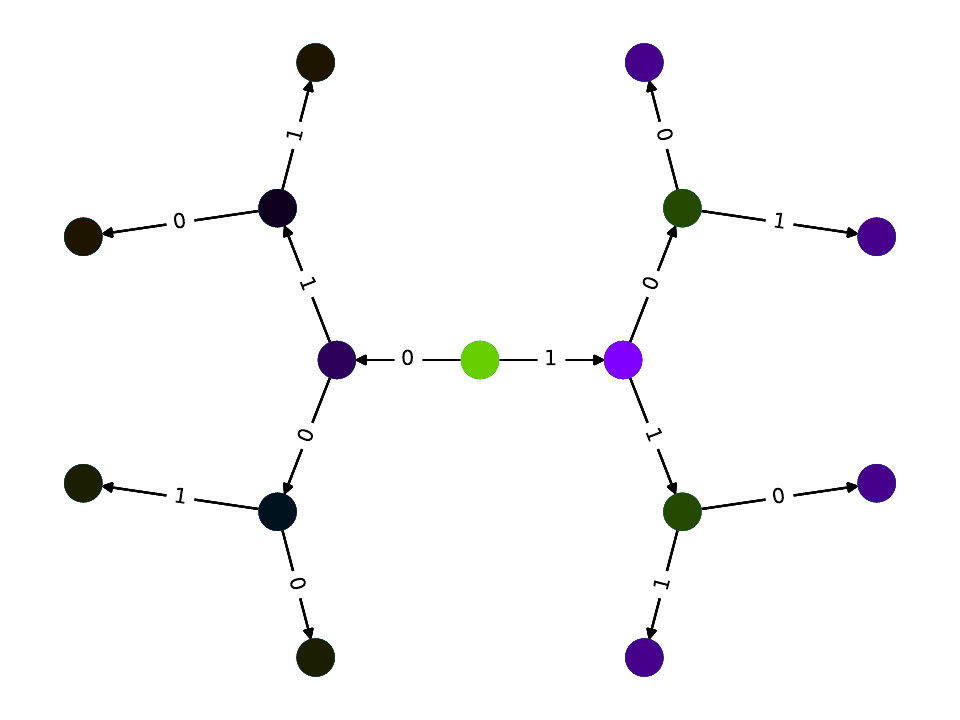}
        \subcaption{$R_AR_BR_A\ket{r}$}
        \label{fig:TreeDepthThree_2A}
    \end{subfigure}
    \begin{subfigure}[]{0.5\textwidth}
        \centering
        \includegraphics[scale=0.4,trim={0.5cm 0cm 0cm 0cm}]{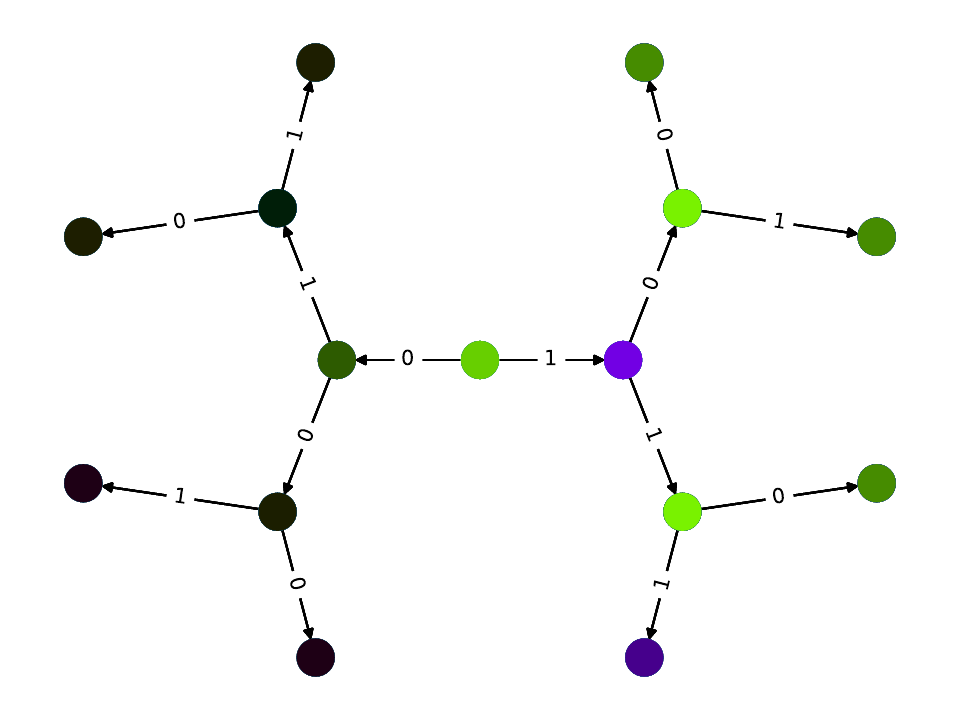}
        \subcaption{$(R_BR_A)^2\ket{r}$}
        \label{fig:TreeDepthThree_2B}
    \end{subfigure}
\caption{Visualisation of a depth 3 \texttt{QuantumBacktrackingTree}, initialized in the state $\ket{r}$, for successive applications of the operators $R_A$ and $R_B$. The colors of the nodes (green, purple) indicate the sign of their non-zero amplitudes in the node state $\ket{x}$.
Fig.~\ref{fig:TreeDepthThree_1A}: $\ket{x}=R_A\ket{r}$. 
Fig.~\ref{fig:TreeDepthThree_1B}: $\ket{x}=R_BR_A\ket{r}$. The branches originating from the rejected node [0] are not explored, as the diffuser $R_B$ acts as $-\I$ on the associated node state $\ket{x}=\ket{[0,0,0]}\ket{2}$.  
Fig.~\ref{fig:TreeDepthThree_2A}: $\ket{x}=R_AR_BR_A\ket{r}$. 
Fig.~\ref{fig:TreeDepthThree_2B}: $\ket{x}=(R_BR_A)^2\ket{r}$.
}
\label{fig:TreeDepthThree_qstep}
\end{figure*}

Subsequently, we successively apply the diffusion operators $R_A$ and $R_B$ constituting one \texttt{quantum\_step}, as shown in the listings below. This step is repeated twice with the belonging backtracking trees illustrated in Figure \ref{fig:TreeDepthThree_qstep}. 
With each application of the diffuser, the quantum walk explores a deeper layer of the tree.
Note that the branches originating from the rejected node with path [0] from the root are not explored by the quantum walk. 

\begin{lstlisting}[language = Python, numbers = none]
# Apply R_A
tree.qstep_diffuser(even=False)
tree.visualize_statevector()

# Apply R_B
tree.qstep_diffuser(even=True)
tree.visualize_statevector()
\end{lstlisting}

\subsection{The \texttt{quantum\_step} function}\label{QuantumStepFunction}

This function applies the operator $R_BR_A$ on the \texttt{QuantumBacktrackingTree} instance. To do so, it employs the \texttt{qstep\_diffuser} with the parameter $\vareven$ twice. $\vareven$ refers to the parity of the height attribute instead of the distance from the root.
Depending on the parameter $\vareven$, the \texttt{qstep\_diffuser} acts on the subspaces $\mathcal H_x$ where $x$ has odd ($\vareven=\texttt{False}$) or even ($\vareven=\texttt{True}$) height.
If the $\varmaxdepth$ of the tree is odd, then $\vareven=\texttt{False}$ corresponds to $R_A$ (otherwise $R_B$), and vice versa if the $\varmaxdepth$ is even. 


\subsubsection{The \texttt{qstep\_diffuser} function}

This function applies the operator $R_A=\bigoplus_{x\in A} D_x$ or $R_B=\ket{r}\bra{r}+\bigoplus_{x\in B} D_x$ on the \texttt{QuantumBacktrackingTree} instance. As explained above, it receives the parameter $\vareven$ which specifies whether it acts on the subspaces $\mathcal H_x$ where $x$ has odd or even height. Depending on the $\varmaxdepth$ of the tree this then corresponds to $R_A$ or $R_B$.

Let $U_x$ be a unitary such that $\ket{\psi_x}=U_x\ket{x}$. The unitaries $U_A=\bigoplus_{x\in A} U_x$ and $U_B=\bigoplus_{x\in B} U_x$ are implemented as shown in Algorithm \ref{alg:psi_prep}. In Montanaro's algorithm rejected nodes are never explored as they simply do not appear as children of their parents. Our implementation, however, explores the rejected nodes while also ensuring they have 0 children: For implementing the unitary $U_x$, we do not check whether a node $x$ is rejected. Thus, $U_x\ket{x}$ yields a superposition of the state $\ket{x}$ and all of its possible child states $\ket{y}$ in a full tree of depth $N=\varmaxdepth$. To ensure that $D_x$ acts as $D_x=-\I$ on a rejected node $x$, we therefore have to flip the phase of the state $\ket{x}$ and all such child states $\ket{y}$ if $x$ is rejected.

The operator $D_x$ acts as $D_x=\I$ if $x$ is marked, $D_x=-\I$ if $x$ is rejected, and $D_x=U_x(\I-2\ket{x}\bra{x})U_x^{\dagger}$ otherwise. This can be expressed as
\begin{equation}
D_x=U_xO_1(x)O_2(x)U_x^{\dagger},
\end{equation}
where 
\begin{equation}
\label{O1}
O_1(x)=\I-(1+(-1)^{\accept(x)})\ket{x}\bra{x},
\end{equation}
and
\begin{equation}
\label{O2}
O_2(x)=\I-(1-(-1)^{\reject(x)})\sum\limits_{y,x\rightarrow x}\ket{y}\bra{y}.
\end{equation}

The operator $O_1(x)$ acts as the identity if the node $x$ is accepted.
On the other hand, if $x$ is not accepted, it flips the phase of the node state $\ket{x}$.

The operator $O_2(x)$ acts as the identity if the node $x$ is not rejected.
It flips the phase of each child state in $\{\ket{y}\mid x\rightarrow y\}$ if $x$ is rejected. In this case, the operator $O_1(x)O_2(x)$ flips the phase of the state $\ket{x}$, together with the phase of each child state in $\{\ket{y}\mid x\rightarrow y\}$, so that $D_x=U_x(-\I)U_x^{\dagger}=-\I$.

For the implementation of the operators we use the so-called $\oddity$ function that returns a \texttt{QuantumBool} when called on a node. Depending on whether we perform the diffusion operators $D_x$ on the subspaces $\mathcal H_x$ such that $x$ has odd (even) height, $\oddity(x)$ is \texttt{True} if and only if the height of $x$ is odd (even). Then, if $\oddity(x)$ is \texttt{True}, $\oddity(y)$ is \texttt{False} for all children $y$ of $x$.

The operator $O_1(x)$ is implemented as 
\begin{equation}
\label{O1_CZ}
\CZ\big(\nnot \accept(x),\oddity(x)\big).
\end{equation}
Controlling the operator on $\oddity(x)$ ensures that we do not flip the phase of a child state in $\{\ket{y}\mid x\rightarrow y\}$ if $y$ is also not accepted.

The operator $O_2(x)$ is implemented as 
\begin{equation}
\label{O2_CZ}
\CZ\big(\reject(\hat{x}),\nnot \oddity(x)\big)
\end{equation}
where $\hat{x}$ denotes the parent of node $x$. The operation that maps the child states to their parent is called \textit{lifting}. Similarly, controlling the operator\footnote{"Controlling the operator $O$" is jargon and means, compiling the controlled version $cO$. For more information on controlling consider section \ref{sec:custom_control}.} on the negation of $\oddity(x)$ ensures that it only operates non-trivially on the child states. Controlling the operator on $x$ not being the root ensures that $O_2(x)$ acts as identity on the root. 

We now describe the process of applying the operator $O_2(x)$, which applies a phase to the child states of node $x$ if the reject function on $x$ returns true. This raises the problem of applying a phase to the child states $\{ \ket{y} \mid x\rightarrow y\}$, using information only obtainable from the parent state $\ket{x}$. To achieve this, we perform an operation that we call \textit{lifting}. The lifting operation temporarily maps every child state to its parent. We then evaluate the reject function on the lifted state and apply the phase using the $\MCZ$ gate, before reverting the lifting. In the following the technical details of the lifting operation are described.

Lifting a state $\ket{y}$ of height $i$ to its parent $\ket{x}$ of height $i+1$ works by first incrementing the height variable $\varh$. This can be done without any quantum gates because $\varh$ is one-hot encoded. The increment is just a compiler swap: The last qubit of $\varh$ is moved to the front.\footnote{If $\ket{x}$ is the root this is a special case which has to be treated differently: It is mapped to a leaf by this lifting procedure. Note that this only becomes a problem if the parity of the maximum depth $N$ of the tree is equal to the $\vareven$ keyword of the \texttt{qstep\_diffuser}. In this case, we can prevent the behavior by applying the gate $\CX(\varh[N], \oddity(x))$. This flips the state of the from $\ket{1}$ to $\ket{0}$, and therefore prevents the erroneous execution of the $\MCZ$ gate.}

This could potentially leave us in a non-algorithmic state. To understand why this might be the case consider the following example:
\begin{equation}
\begin{aligned}
    \ket{\texttt{h}}\ket{\texttt{branch\_qa}}
    = & \ket{2}\ket{[0,0,1,1,1]}\\
    \rightarrow & \ket{3}\ket{[0,0,1,1,1]}
\end{aligned}
\end{equation}
The first "1" in $\varbranchqa$ is now at a position of height 2, even though the $\varh$ variable says, the height is 3. To remedy this problem, we create a temporary container with the quantum data type of the entries of $\varbranchqa$. We now perform a series of swaps into the temporary container variable with every potential entry of $\varbranchqa$ that could exhibit this type of problem. In order to modify only the relevant entry, each swap for entry $i$ is controlled on the qubit $\varh[i]$.

Since controlled swap gates can be rather costly, we exhibit the \texttt{subspace\_optimization} keyword for the constructor of the \texttt{QuantumBacktrackingTree} class. This keyword allows the user to indicate that the \texttt{reject} function is guaranteed to return the same results on the non-algorithmic subspace of each node. This is especially the case for the \texttt{reject} function describing Sudoku instances. Therefore, if this keyword is set to \texttt{True}, the lifting operation consists only of the compiler swap and requires a grand total of 0 quantum gates.\\

As the final piece of the implementation, the \texttt{qstep\_diffuser} is described in Algorithm~\ref{alg:qstep_diffuser}\footnote{The line \texttt{with invert} indicates that the inverse of the quantum instruction in the indented area is applied. Similarly, the line \texttt{with control arg do} indicates that every quantum instruction in the indented area is controlled on the qubit \texttt{arg}.}.
Note that all auxiliary quantum variables, e.g., $\varreject$, $\varaccept$, $\varoddity$, have to be uncomputed. In \textit{Qrisp}, 
this is handled automatically by the uncomputation algorithm \cite{Seidel_2023_uncomputation} which is called with the \texttt{auto\_uncompute} decorator.

Additionally, this function may receive a list of qubits $\varctrl$. 
This is beneficial because this operator will undergo phase estimation and we want to prevent performing automatic synthesis of the controlled operation,
as this would imply controlling every gate. Instead we can just control the $\MCZ$ gates. This is further explained in Appendix \ref{appendix:qstep}.

\begin{algorithm}[]
\caption{\texttt{qstep\_diffuser}}\label{alg:qstep_diffuser}

\KwIn{
  \begin{itemize}
    \item A \texttt{QuantumBacktrackingTree} specified by $N=\varmaxdepth$, a superposition of node states 
    $\ket{x}=\ket{\varbranchqa}\ket{\varh}$, and predicate functions $\accept$, $\reject$.
    \item A Boolean $\vareven$.
    \item A list of control qubits $\varctrl$.
  \end{itemize}
}

\tcp{Perform $U_x^\dagger$}
\Invert{
    \texttt{psi\_prep}$(x, N, \vareven)$\;
}

\tcp{Compute \texttt{oddity} function}
$\varoddity \gets \texttt{False}$\;
\For{$i = 0$ \KwTo $N$}{
    \If{$i \bmod 2 \neq \vareven$}{
        \KwCX $\varh[i]$, $\varoddity$\;
    }
}

\tcp{Compute \texttt{accept} function}
$\varaccept \gets \accept(x)$\;

\tcp{Perform the operator $O_1(x)$}
\KwMCZ{
	$\nnot \varaccept$, $\varoddity$, $\varctrl$\;
}

\tcp{Lifting: states $x$ are mapped to their parent $\hat{x}$}
\tcp{Special case for root}
\If{$N \bmod 2 == \vareven$}{
    \KwCX $\varh[N]$, $\varoddity$\;
}

\tcp{Swap branch information into a temporary container}
$\mathrm{temp\_qv} \gets 0$\;
\For{$i = 0$ \KwTo $N-1$}{
    \If{$i \bmod 2 == \vareven$}{
        \Control{$\varh[i]$}{
            \KwSwap $\vartempqv$, $\varbranchqa[i]$\;
        }
    }
}

\tcp{Increment height}
compiler swap: remove qubit $\varh[N]$ and insert it at first place of $\varh$\;

\tcp{Compute \texttt{reject} function}
$\varreject \gets \reject(x)$\;

\tcp{Perform the operator $O_2(x)$}
\KwMCZ{
	$\varreject$, $\nnot \varoddity$, $\varctrl$\;
}

\tcp{Decrement height}
reverse compiler swap: remove qubit $\varh[0]$ and insert it at last place of $\varh$\;

\tcp{Reintroduce branching information; uncomputes the temporary container}
\For{$i = 0$ \KwTo $N-1$}{
    \If{$i \bmod 2 == \vareven$}{
        \Control{$\varh[i]$}{
            \KwSwap $\vartempqv$, $\varbranchqa[i]$\;
        }
    }
}

\tcp{Delete temporary container}
delete $\vartempqv$\;

\tcp{Perform $U_x$}
\texttt{psi\_prep}$(x, N, \vareven)$\;
    
\end{algorithm}

%% file: sections/psi_prep.tex
\subsection{The \texttt{psi\_prep} function}
To perform the \texttt{qstep\_diffuser} function shown in Algorithm \ref{alg:qstep_diffuser}, we must first implement an auxiliary function realizing the operator $U_x$ with the property $U_x\ket{x}=\ket{\psi_x}$. Such a function \texttt{psi\_prep}, when acting on state $\ket{x}$, results in the state $\ket{\psi_x}$.
The function \texttt{psi\_prep} receives a parameter $\vareven$, which specifies whether it acts on the subspaces $\mathcal H_x$ where $x$ has odd or even height.

The main ideas for implementing $U_x$  are summarized in the following two steps:
\begin{itemize}
    \item 
First, we manipulate the height $\ket{\varh}=\ket{i}$ of state $\ket{x}$ by utilizing a continuous swap resulting in a superposition of heights $\ket{i}$ and $\ket{i-1}$, that is,  $\ket{i}\rightarrow\ket{i}+\gamma\ket{i-1}$
for some value $\gamma$ to be specified.

\item
Secondly, we set up a superposition of state $\ket{x}$ at height $i$ and its child states $\ket{y}$ at height $i-1$. This is achieved by applying a Hadamard gate to the \texttt{QuantumVariable} $\varbranchqa[i-1]$ controlled on the qubit $\varh[i-1]$.
\end{itemize}

With these steps, the unitary $U_x$, as implemented in Algorithm \ref{alg:psi_prep}, acts on a state $\ket{x}$ of height $i$ with child states $\{\ket{y}\mid x\rightarrow y\}$ of height $i-1$ as
\begin{equation}
\ket{\psi_x}\propto\ket{x}+c\sum_{y,\, x\rightarrow y}\ket{y}
\end{equation}
where $c=\sqrt{N}$ if $x$ is the root and $c=1$ otherwise.

Each call of \texttt{psi\_prep} therefore, when applied to a parent state $\ket{x}$, prepares the superposition state $\ket{\psi_x}$ of the parent state and its child states $\ket{y}$, and does that for every subspace $\mathcal{H}_x$. 

Let us examine how \texttt{psi\_prep} acts on state $\ket{x}=\ket{[0,0,0,1]}\ket{3}$ in a binary tree of depth $4$: In this case, $\ket{x}$ has two children, $\ket{y_0}$ and $\ket{y_1}$, and the resulting state is 
\begin{align}
\begin{split}
\ket{\psi_x}=U_x\ket{x}=&\frac{1}{\sqrt{3}}\big(\ket{x}+\ket{y_0}+\ket{y_1}\big)\\
=&\frac{1}{\sqrt{3}}\big(\ket{[0,0,0,1]}\ket{3}\\
&\hspace{3,5mm}+\ket{[0,0,0,1]}\ket{2}\\
&\hspace{3,5mm}+\ket{[0,0,1,1]}\ket{2}\big).
\end{split}
\end{align}

\begin{figure}[H]
    \hspace*{0.4cm}
    \begin{subfigure}[]{0.22\textwidth}
        \centering
        \includegraphics[clip, trim=0.275cm 0cm 0cm 0cm, scale=0.55]{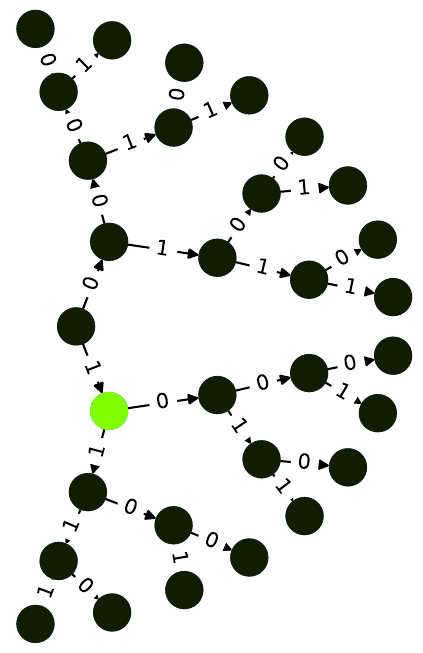}
        \subcaption{}
        \label{fig:before}
    \end{subfigure}
    \begin{subfigure}[]{0.22\textwidth}
        \centering
        \includegraphics[clip, trim=0.195cm 0cm 0cm 0cm 0cm, scale=0.55]{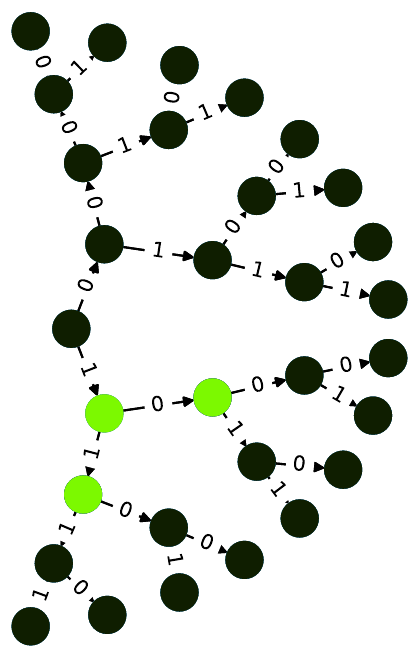}
        \subcaption{}
        \label{fig:after}
    \end{subfigure}

\hspace*{-0.45cm}
\begin{subfigure}[]{0.40\textwidth}
\centering
\scalebox{0.75}{
\Qcircuit @C=1.0em @R=0.2em @!R { \\
	 	\nghost{{b\_3.0} :  } & \lstick{{b\_3.0} :  } & \gate{\mathrm{X}} & \qw & \qw & \qw & \qw & \qw & \qw\\
	 	\nghost{{b\_2.0} :  } & \lstick{{b\_2.0} :  } & \qw & \qw & \ctrlo{4} & \qw & \gate{\mathrm{H}} & \qw & \qw\\
	 	\nghost{{b\_1.0} :  } & \lstick{{b\_1.0} :  } & \qw & \qw & \qw & \qw & \qw & \qw & \qw\\
	 	\nghost{{b\_0.0} :  } & \lstick{{b\_0.0} :  } & \qw & \ctrlo{4} & \qw & \gate{\mathrm{H}} & \qw & \qw & \qw\\
	 	\nghost{{h.4} :  } & \lstick{{h.4} :  } & \qw & \qw & \qw & \qw & \qw & \qw & \qw\\
	 	\nghost{{h.3} :  } & \lstick{{h.3} :  } & \gate{\mathrm{X}} & \qw & \multigate{1}{\mathrm{(XX+YY)}\,({\ensuremath{\Phi}})}_<<<{1} & \qw & \qw & \qw & \qw\\
	 	\nghost{{h.2} :  } & \lstick{{h.2} :  } & \qw & \qw & \ghost{\mathrm{(XX+YY)}\,({\ensuremath{\Phi}})}_<<<{0} & \qw & \ctrl{-5} & \qw & \qw\\
	 	\nghost{{h.1} :  } & \lstick{{h.1} :  } & \qw & \multigate{1}{\mathrm{(XX+YY)}\,({\ensuremath{\Phi}})}_<<<{1} & \qw & \qw & \qw & \qw & \qw\\
	 	\nghost{{h.0} :  } & \lstick{{h.0} :  } & \qw & \ghost{\mathrm{(XX+YY)}\,({\ensuremath{\Phi}})}_<<<{0} & \qw & \ctrl{-5} & \qw & \qw & \qw\\
\\ }}
\vspace*{-0.3cm}
\subcaption{}
\label{fig:cXXYYswap}
\end{subfigure}
\caption{Visualisation of a depth 4 \texttt{QuantumBacktrackingTree}. 
Fig.~\ref{fig:before}: The state $\ket{x}=\ket{[0,0,0,1]}\ket{3}$ is initialized. 
Fig.~\ref{fig:after}: After applying \texttt{psi\_prep}, we can observe the resulting state $\ket{\psi_x}$ as a superposition of $\ket{x}$ and its child states $\ket{y_0}=\ket{[0,0,0,1]}\ket{2}$, and $\ket{y_1}=\ket{[0,0,1,1]}\ket{2}$.\\
Fig.~\ref{fig:cXXYYswap}: The resulting quantum circuit applied by \texttt{psi\_prep}. First the height $\ket{\varh}=\ket{3}$ and $\ket{b}=\ket{\varbranchqa}=\ket{[0,0,0,1]}$ are initialized. Then, two controlled $\XXYY$ gates (with $\beta=\pi/2$) are applied to maneuver the position of bit $1$ into a superposition of $\ket{3}\rightarrow\ket{3}+\ket{2}$. It is worth mentioning that one should pay attention to the knob of the controlled gate - the circle denotes that, contrary to a regular controlled gate, this one activates if the control qubit is in state $\ket{0}$. Lastly, two controlled Hadamard gates are applied in order to set up the superposition in $\varbranchqa$.
}
\label{fig:psi_prep_all}
\end{figure}

A visualisation of the tree in question initialized in state $\ket{x}$ before (Fig.~\ref{fig:before}), and after (Fig.~\ref{fig:after}) applying $\texttt{psi\_prep}$, together with the quantum circuit of $\texttt{psi\_prep}$ for this exact example (Fig.~\ref{fig:cXXYYswap}) is provided in Fig.~\ref{fig:psi_prep_all}.

In order to manipulate the height of the state we generate a controlled $\XXYY$ gate\footnote{Refer to Appendix \ref{sec:custom_control_xxyy} for more details.}, which acts similarly to a parameterized swap. It allows us to control the position of the $1$ bit of the one-hot encoded integer and maneuver it to where we want it to be utilizing said swaps, e.g., $\ket{3}=\ket{0001}\rightarrow\ket{0001}+\ket{0010}=\ket{3}+\ket{2}$. The unitary matrix of the $\XXYY$ gate is 
\begin{equation*}
R_{\XXYY}(\Phi,\beta=\frac{\pi}{2})=
\begin{pmatrix}
1&0&0&0\\
0&\cos{\frac{\Phi}{2}}&-\sin{\frac{\Phi}{2}}&0\\
0&\sin{\frac{\Phi}{2}}&\cos{\frac{\Phi}{2}}&0\\
0&0&0&1
\end{pmatrix}. 
\end{equation*}
Depending on the rotation angles $\Phi$ and $\beta$, the $\XXYY$ gate can correspond to performing a swap, or acts as an identity if $\Phi=4n\pi$.

After successfully moving the ``one-hot $1$'' we take the second step mentioned above and apply a Hadamard gate controlled on the target of the ``movement'' in order to set up the superposition in $\varbranchqa$.
Thereby, it is important to make certain of the fact that the non-algorithmic states remain invariant under performing $U_x$ in order for them not to get tagged in the $\texttt{qstep\_diffuser}$. We achieve this by controlling the parameterized swap on the \texttt{QuantumVariable} $\varbranchqa[i-1]$ which is set to superposition by the controlled H-gate. This $\texttt{QuantumVariable}$ represents the branch information of the child states. If this is a non-zero state, while the height variable indicates the parent state, the state is indeed invariant.

\begin{algorithm}[]
\caption{\texttt{psi\_prep}}\label{alg:psi_prep}

\KwIn{
  \begin{itemize}
    \item A superposition of node states $\ket{x}=\ket{\varbranchqa}\ket{\varh}$ and $N=\varmaxdepth$ for a 
    \texttt{QuantumBacktrackingTree}.
    \item A Boolean $\vareven$.
  \end{itemize}
}

$root\_phi \gets 2\arctan(\sqrt{N\cdot \vardegree})$\;
$phi \gets 2\arctan(\sqrt{\vardegree})$\;
\tcp{Node $x$ is root}
\If{$N \bmod 2 \neq \vareven$}{
    \Control{$\nnot \varbranchqa[N-1]$}{
        \KwXXYY $root\_phi$, $\varh[N-1]$, $\varh[N]$\;
    }
    \Control{$\varh[N-1]$}{
        \KwH $\varbranchqa[N-1]$\;
    }
}
\tcp{Node $x$ is not root}
\For{$i = \vareven$ \KwTo $N-2$ \KwBy $2$}{
    \Control{$\nnot \varbranchqa[i]$}{
        \KwXXYY $phi$, $\varh[i]$, $\varh[i+1]$\;
    }
    \Control{$\varh[i]$}{
        \KwH $\varbranchqa[i]$\;
    }
}
\end{algorithm}

Note that the \texttt{prep\_psi} function essentially treats all parent/child subspaces $\mathcal H_x$ in parallel. This implies that this function has constant depth. If the \texttt{subspace\_optimization} keyword is set to \texttt{True}, the \texttt{qstep\_diffuser} function therefore consists of two additional 2-controlled Z gates (also constant depth) and the computation of the oddity which scales very weakly linearly with the maximum height of the tree. Therefore, the \texttt{qstep\_diffuser} has \textit{almost} constant quantum circuit depth.

%% file: sections/oracle_implementation.tex
\section{Oracle implementation}
\label{sec:oracle_implementation}

Similar to Grover's Algorithm \cite{grover}, Montanaro's backtracking algorithm revolves around the evaluation of unspecified black box functions (or \textit{oracles}). In the Grover setting, a solution is marked if it satisfies some (application specific) condition. In the backtracking setting we have a similar situation. However, instead of having only one oracle, we now have two oracles where the first one specifies whether a node satisfies a condition (called \texttt{accept}), and the other one specifies whether the algorithm should no longer consider the subtree of that node (called \texttt{reject}). For the Qrisp implementation, these functions need to take an instance of the \texttt{QuantumBacktrackingTree} class and return a \texttt{QuantumBool}, which indicates whether the node is accepted or rejected.\\
In this section, we describe the implementation of these functions.
\subsection{Sudoku encoding}
The \textit{Qrisp} implementation of Montanaro's algorithm provides the user with a \texttt{QuantumArray branch\_qa} describing the assignments that the backtracking algorithm took so far. This \texttt{QuantumArray} consists of $n$ \texttt{QuantumVariables} of a given quantum type, where $n$ is the maximum depth of the backtracking tree. Given a Sudoku instance with $k \in \mathbb{N}$ empty entries, we model the problem with a tree of depth $n = k+1$ (the $+1$ is elaborated upon below) with \texttt{QuantumVariables} of type \texttt{QuantumFloat}\footnote{\texttt{QuantumFloat} is the quantum type to describe numbers in \textit{Qrisp}. This type supports (signed) integers but also fractional numbers.\cite{Seidel_2022}}, to describe the fact that a Sudoku solution consists of a sequence of integers.

Given the \texttt{accept} and \texttt{reject} functions described below, the code for setting up the \texttt{QuantumBacktrackingTree} could be given as:

\begin{lstlisting}[language = Python, numbers = none]
from qrisp import *
from qrisp.quantum_backtracking
     import QuantumBacktrackingTree as QBT

num_empty_fields = 7

tree = QBT(max_depth = num_empty_fields+1,
           branch_qv = QuantumFloat(2),
           accept = accept,
           reject = reject)
\end{lstlisting}

\subsection{The \texttt{accept} function}

This function is rather simple: A Sudoku board is solved correctly if all entries are filled with numbers that do not contradict the rules of Sudoku. In \textit{backtracking language} this means, that a node is accepted if it has height $0$ and none of its ancestor nodes were rejected. Thus, the implementation of this function is rather simple:

\begin{lstlisting}[language = Python, numbers = none]
@auto_uncompute    
def accept(tree):
    return tree.h == 0
\end{lstlisting}

However, there is a caveat for practical reasons: While Montanaro suggests that the algorithm should never explore rejected nodes, in our implementation rejected nodes are explored but have no children. As described above, we need to pick the depth to be $n = k + 1$ where $k$ is the number of empty fields in the Sudoku board. Otherwise, i.e., if $n = k$, the sibling nodes of the solution might be rejected. Because of this fact, the algorithm will still explore them and evaluate \texttt{accept} to \texttt{True} (because they have height 0), leading to the ambiguous situation that a node returns \texttt{True} for both \texttt{reject} and \texttt{accept}.

\subsection{The \texttt{reject} function}
The \texttt{reject} function is more complicated because this function needs to consider the Sudoku board and check whether all the assignments are in compliance with the rules of Sudoku. Another layer of complexity is introduced by the fact that the \texttt{reject} function should only consider entries that have already been assigned. To keep our presentation comprehensive, we will first describe how a fully assigned Sudoku board can be checked, and modify this function such that it can also ignore non-assigned values afterwards.

\subsubsection{Mapping to a graph-coloring problem}
To check the compliance of a fully assigned Sudoku board (encoded in \texttt{branch\_qa}), the first step is to transform it into a graph-coloring problem. This implies that we represent each entry of the Sudoku board (given or assigned) as a node of an undirected graph $G$. The rules of Sudoku (columns, rows, and squares containing only distinct entries) are then included by adding an edge to $G$ for each comparison that needs to be performed to assert distinctness of the elements.

\begin{algorithm}[]
\caption{Graph coloring problem from Sudoku instance}\label{alg:g_coloring}

\KwIn{An $N^2 \times N^2$ array $A$ representing a Sudoku board, where the empty fields are marked by -1.}
\KwOut{A graph $G$ such that a coloring (respecting the predetermined Sudoku values) would be a solution of the Sudoku board.}

$G \gets \text{Graph}()$\;
\For{$i \gets 0$ \KwTo $N^2-1$}{
    \For{$j \gets 0$ \KwTo $N^2-1$}{
        \eIf{$A[i,j] == -1$}{
            $node\_type \gets \text{assigned}$\;
        }{
            $node\_type \gets \text{given}$\;
        }
        $G.\text{add\_node}((i,j), \text{node\_type} = node\_type)$\;
    }
}

\For{$i \gets 0$ \KwTo $N^2-1$}{
    \For{$j \gets 0$ \KwTo $N^2-1$}{
        \If{$A[i,j] == -1$}{
            \tcp{Add column checks}
            \For{$k \gets 0$ \KwTo $N^2-1$}{
                \If{$k == j$}{
                    \KwContinue\;
                }
                \eIf{$A[i,k] == -1$}{
                    $edge\_type \gets \text{qq}$\;
                }{
                    $edge\_type \gets \text{cq}$\;
                }
                $G.\text{add\_edge}((i,j),(i,k), \text{type} = edge\_type)$\;
            }
            \tcp{Row and square checks work the same}
        }
    }
}

\KwRet $G$\;
\end{algorithm}

For obvious reasons, we add an edge only if at least one of the participating nodes represents an assigned field. Furthermore, we distinguish between \textit{quantum-quantum} edges, i.e., a comparison between two empty fields,  and \textit{classical-quantum} edges. This is because for any given node the latter type can be batched together into a single quantum logic synthesis \cite{soeken2017logic, Meuli2019,Porwik2002, Amy_2019, Seidel_2023} call (discussed in the next section).

\subsubsection{Evaluating the comparisons}
The next step is to evaluate the comparisons to check for element distinctness. This means that we iterate over the edges of the graph and compute a \texttt{QuantumBool} for each edge indicating distinctness.
For this we distinguish between the \textit{quantum-quantum} and the \textit{classical-quantum} comparison cases. For the first case we simply call the $==$ operator on the two participating quantum variables to compute the comparison \texttt{QuantumBool}. The circuit employed by this function can be found in Appendix \ref{sec:controled_qq_comparison}.

As mentioned earlier, classical-quantum comparisons can be batched together to be evaluated in a single function call. In \textit{Qrisp} this is performed using the \texttt{QuantumDictionary} class\footnote{\textit{Qrisp}'s \texttt{QuantumDictionary} uses logic synthesis for evaluation. For our implementation we used Gray-synthesis \cite{Amy_2019} but other approaches are also possible, for instance, PPRM \cite{Porwik2002} or pebble-games \cite{Meuli2019}.}:

\begin{lstlisting}[language = Python, numbers = none]
def cq_eq_check(q_value, cl_values):
    """
    Receives a QuantumVariable and a list of classical
    values and returns a QuantumBool, indicating whether
    the value of the QuantumVariable is contained in the
    list of classical values
    """

    # Create dictionary
    qd = QuantumDictionary(return_type = QuantumBool())

    # Determine the values that q_value can assume
    value_range = [q_value.decoder(i) for i in range(2**q_value.size)]
    
    # Fill dictionary with entries
    for value in value_range:
        if value in cl_values:
            qd[value] = True
        else:
            qd[value] = False

    # Evaluate dictionary with quantum value
    return qd[q_value]
\end{lstlisting}

This function receives a \texttt{QuantumVariable} and a list of classical values and returns a \texttt{QuantumBool} which indicates whether the quantum value can be found in the list of the classical values.
As an example consider \texttt{cl\_values = [1,2,3]}. If we then call this function (represented by the unitary $U$) on a \texttt{QuantumVariable} in superposition of 0 and 3, we get
\begin{align}
U(\ket{0} + \ket{3})\ket{\text{False}} = \ket{0}\ket{\text{False}} + \ket{3}\ket{\text{True}},
\end{align}
because 0 is not in the given list but, 3 is.

Putting everything together yields a list of \texttt{QuantumBool}, where each entry indicates whether the comparison result is \texttt{True}. Since the rules of Sudoku require element distinctness, the assignment is therefore valid, if every entry of the list is in the \texttt{False} state. We can check this by simply allocating another \texttt{QuantumBool} and executing a multi-controlled X-gate (with control state 0) on the list of comparison results. Since this list can be extensive, we utilize the logarithmic depth MCX implementation given in \cite{Balauca_2022}.

Finally, the last step is to uncompute every temporary variable (such as the comparison results). Fortunately \textit{Qrisp} automates this procedure using the \texttt{auto\_uncompute} decorator \cite{Seidel_2023_uncomputation, Paradis_2021}, so that we do not need to care about this explicitly.

\subsubsection{Ignoring non-assigned fields}
As this is a backtracking implementation, our Sudoku compliance check also has to understand that the results of certain comparisons should be ignored, since the corresponding fields are not assigned yet. For example, consider a Sudoku field with 4 empty fields, where only one field has been assigned so far. In our implementation of the algorithm, the empty fields are encoded as zeros in \texttt{branch\_qa} and we only know that they are not assigned yet by considering the height \texttt{QuantumVariable h}. The implementation of the Sudoku-check algorithm given above would therefore return \textit{not valid} for almost every single node, because it assumes that the 3 remaining empty fields carry the value 0 even though in reality they have not been assigned yet. We therefore need to consider the value of \texttt{h} accordingly.

Fortunately, the one-hot encoding of this variable makes this rather easy: The value that has been assigned most recently is indicated by the corresponding qubit in \texttt{h} being in the $\ket{1}$ state. For example, in a tree of maximum depth 5, if the \texttt{branch\_qa} entry with height 3 has been assigned recently, \texttt{h} will be in the state $000100$. The next assignment would then be height 2, i.e. $001000$.
For a quantum-classical comparison with the \texttt{branch\_qa} entry $i$, we can therefore simply call the comparison evaluation controlled on the $i$-th qubit in \texttt{h}. This implies that this comparison can only result in \texttt{True}, and as a result cause the \texttt{reject} value to be \texttt{True} if $i$ was assigned most recently.\\
\newpage
\begin{lstlisting}[language = Python, numbers = none]
def eval_cq_checks( batched_cq_checks, 
                    q_assigments, 
                    h):
    """
    Batched cq_checks is a dictionary of the form
    
    {int : list[int]}
    
    Where each key/value pair corresponds to 
    one batched quantum-classical comparison.
    The keys represent the the quantum values 
    as indices of q_assigments and the values
    are the list of classical valuesthat 
    the quantum value should be compared with.
    q_assigments is a QuantumArray derived from
    branch_qa and contains the assignments the
    algorithm took so far.
    h is the one-hot encoded QuantumVariable
    that specifies the height of the node
    and specifies which assignment
    should be checked.
    """
    # Create result list
    res_qbls = []

    # Iterate over all key/value pairs to evaluate
    # the comparisons.
    for key, value in batched_cq_checks.items():
        # Enter the control environment
        with control(h[key]):
            # Evaluate the comparison
            eq_qbl = cq_eq_check(q_assignments[key], 
                                 value)
        res_qbls.append(eq_qbl)

    # Return results
    return res_qbls
\end{lstlisting}

The code examples above demonstrate a function that takes a dictionary representing the batched quantum-classical equality checks, the \texttt{QuantumArray q\_assignments}, and the \texttt{QuantumVariable h} as input. It returns a list of of \texttt{QuantumBool} that represent the result of the comparisons. Note the line \texttt{with control(h[key]):} which enters a \texttt{ControlEnvironment}. This means that every quantum instruction that happens in the indented area is controlled on the qubit \texttt{h[key]}. As described above, this feature ensures that the comparison of values that are not assigned yet cannot contribute to the result of the \texttt{reject} function.

We adopt a similar approach for the quantum-quantum comparison. For a comparison between the $i$-th and $j$-th position, we control the comparison on the $k$-th qubit of the \texttt{h} variable where $k = \text{min}(i,j)$. Therefore the value $k$ is the position of the more recently assigned variable. To verify that this implies indeed the desired behavior, we have to distinguish five possible cases accordingly. We assume $i>j$ (otherwise rename the indices), implying the control qubit will always be \texttt{h[j]}:

\begin{enumerate}
    \item Position $i$ and $j$ are assigned but not recently.\\ $\Rightarrow$ No comparison is performed because the control qubit is not in the $\ket{1}$ state.
    \item Position $i$ has been assigned earlier and position $j$ has been recently assigned \\$\Rightarrow$ The comparison is performed because the control qubit is in the $\ket{1}$ state.
    \item Position $i$ has been assigned earlier but position $j$ has not. \\$\Rightarrow$ No comparison is performed because the control qubit is not in the $\ket{1}$ state.
    \item Position $i$ has been recently assigned but position $j$ has not. \\$\Rightarrow$ No comparison is performed because the control qubit is not in the $\ket{1}$ state.
    \item Neither position $i$ nor $j$ are assigned. \\$\Rightarrow$ No comparison is performed because the control qubit is not in the $\ket{1}$ state.

\end{enumerate}

Each of these cases has the desired behavior: The \texttt{reject} function returns \texttt{True} if and only if a newly assigned variable returns \texttt{True} for one of its comparisons.

\begin{lstlisting}[language = Python, numbers = none]
def eval_qq_checks( qq_checks, 
                    q_assignments, 
                    h):
    """
    Batched cq_checks is a list of the form

    [(int, int)]
    
    Where each tuple entry corresponds the index
    of the quantum value that should be compared.
    q_assignments and height are the quantum values
    that specify the tree state.
    """
    # Create result list
    res_qbls = []

    # Iterate over all comparison tuples 
    # to evaluate the comparisons.
    for ind_0, ind_1 in qq_checks:
        # Enter the control environment
        with control(h[min(ind_0, ind_1)]):
            # Evaluate the comparison
            eq_qbl = (q_assignments[ind_0] ==
                      q_assignments[ind_1])
        res_qbls.append(eq_qbl)

    # Return results
    return res_qbls
\end{lstlisting}
This function works similar to \texttt{eval\_cq\_checks}: The comparisons are executed within a control environment and each result is appended to a list of \texttt{QuantumBool}. For an example of how such comparisons work please refer to Fig.~\ref{fig:comparison_tbl}.

\begin{figure*}[t]
    \begin{subfigure}[]{0.5\textwidth}
        \centering
        \begin{tabular}{|c|c|c|c|}
        \hline
        \texttt{q\_assignments} & \texttt{h} & Control & Operands\\
        \hline
        \hline
        3 & 0 & × & \checkmark\\
        \hline
        0 & 0 & \checkmark & \checkmark\\
        \hline
        3 & 1 & × & ×\\
        \hline
        0 & 0 & × & ×\\
        \hline
        0 & 0 & × &×\\
        \hline
        0 & 0 & × &×\\
        \hline
        \multicolumn{2}{|c|}{Comparison Value:}&
        \multicolumn{2}{|c|}{False}
        \\
        \hline
        \end{tabular}
        \subcaption{}
        \label{fig:tablea}
    \end{subfigure}
    \begin{subfigure}[]{0.5\textwidth}
        \centering
        \begin{tabular}{|c|c|c|c|}
        \hline
        \texttt{q\_assignments} & \texttt{h} & Control & Operands\\
        \hline
        \hline
        3 & 0 & × & \checkmark\\
        \hline
        0 & 0 & × & ×\\
        \hline
        3 & 1 & \checkmark & \checkmark\\
        \hline
        0 & 0 & × & ×\\
        \hline
        0 & 0 & × &×\\
        \hline
        0 & 0 & × &×\\
        \hline
        \multicolumn{2}{|c|}{Comparison Value:}&
        \multicolumn{2}{|c|}{True}
        \\
        \hline
        \end{tabular}
        \subcaption{}
        \label{fig:tableb}
    \end{subfigure}

    \begin{subfigure}[]{0.5\textwidth}
        \centering
        \begin{tabular}{|c|c|c|c|}
        \hline
        \texttt{q\_assignments} & \texttt{h} & Control & Operands\\
        \hline
        \hline
        3 & 0 & × & ×\\
        \hline
        0 & 0 & × & \checkmark\\
        \hline
        3 & 1 & × & ×\\
        \hline
        0 & 0 & \checkmark & \checkmark\\
        \hline
        0 & 0 & × &×\\
        \hline
        0 & 0 & × &×\\
        \hline
        \multicolumn{2}{|c|}{Comparison Value:}&
        \multicolumn{2}{|c|}{False}
        \\
        \hline
        \end{tabular}
        \subcaption{}
        \label{fig:tablec}
    \end{subfigure}
    \begin{subfigure}[]{0.5\textwidth}
        \centering
        \begin{tabular}{|c|c|c|c|}
        \hline
        \texttt{q\_assignments} & \texttt{h} & Control & Operands\\
        \hline
        \hline
        3 & 0 & × & ×\\
        \hline
        0 & 0 & × & ×\\
        \hline
        3 & 1 & × & ×\\
        \hline
        0 & 0 & × & \checkmark\\
        \hline
        0 & 0 & × & × \\
        \hline
        0 & 0 & \checkmark &\checkmark\\
        \hline
        \multicolumn{2}{|c|}{Comparison Value:}&
        \multicolumn{2}{|c|}{False}
        \\
        \hline
        \end{tabular}
        \subcaption{}
        \label{fig:tabled}
    \end{subfigure}
    \caption{Examples for four of the five different situations for a quantum-quantum comparison that can appear. The first and second column represent the state of the node (\texttt{h} is one-hot encoded). The path (from the root) is therefore \texttt{[3,0,3]}. The third column indicates on which qubit the comparison evaluation is controlled and the final column indicates which operands participate in the comparison. Note that the control qubit is always at the index of the lower operand. Fig.~\ref{fig:tablea}: Both operands are already assigned but not recently. The comparison has to yield \texttt{False} because otherwise the algorithm would have previously taken a path that is not compliant with the rules of Sudoku. 
    Fig.~\ref{fig:tableb}: Both operands are assigned and one of them recently. The comparison returns \texttt{True} because both \texttt{q\_assignments} entries are in the $\ket{3}$ state. Since one of the operands has been recently assigned, the control qubit is active in this situation. Therefore the comparison will be evaluated and the result is \texttt{True}.
    Fig.~\ref{fig:tablec}: One of the two operands is assigned and the other one is not assigned. The comparison would yield true, because both entries of \texttt{q\_assignments} are in the $\ket{0}$ state. However the comparison is controlled on a qubit of \texttt{h} that is not in the $\ket{1}$ state. Therefore the comparison result is \texttt{False}. 
    Fig.~\ref{fig:tabled}: The comparison would yield True but only because both operands are not assigned yet and therefore carry the value 0. The result of the controlled comparison is still \texttt{False} because the control qubit \texttt{h[0]} is in the $\ket{0}$ state.}
    \label{fig:comparison_tbl}
\end{figure*}

%% file: sections/Experiments.tex
\section{Simulations and Benchmarking}
\label{sec:experiments}

We conduct a series of experiments utilizing the IBM matrix product state (MPS) simulator. Our tests successfully demonstrate the ability to detect and find a solution for 4x4 Sudoku instances with up to 9 empty fields (Figure \ref{fig:sudoku_experiments}). Note that the corresponding circuits can be generated for Sudoku instances with even more empty fields. Yet, as the \texttt{simulator\_mps} allows simulations only for up to 100 qubits, we do not include these instances here. Moreover, since the purpose of this work is to showcase the use of quantum backtracking, we do not apply any classical preprocessing apart from mapping the Sudoku problem to a graph coloring problem. 

\begin{figure}[h]
\centering
\hspace*{-0.5cm}
\includegraphics[scale=0.625,trim={0cm 0cm 0cm 0cm}]{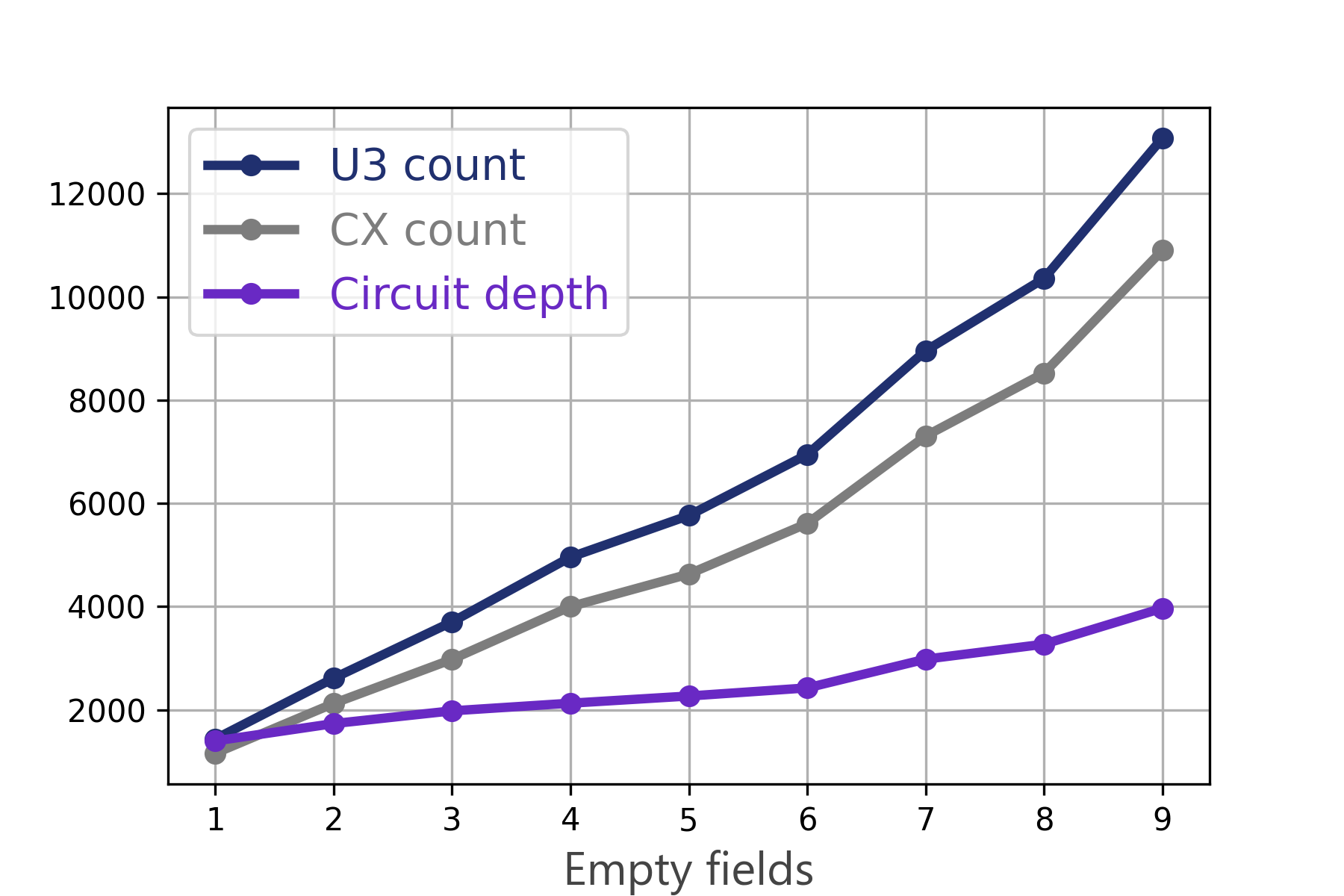}
\caption{Number of basis gates and depth for the circuits implementing quantum phase estimation with precision $2^{-3}$ applied to the operator $R_BR_A$, for Sudoku problems with up to 9 missing cells. The metrics are calculated after transpiling the circuits into the gate set $\{\mathrm{U}3,\mathrm{CX}\}$.}
\label{fig:benchmarking_plot}
\end{figure}

\begin{enumerate}
\item \textit{Detecting a solution.} We benchmark the circuits implementing quantum phase estimation with precision $2^{-3}$ applied to the operator $R_BR_A$ (i.e., the \texttt{quantum\_step}-function) for Sudoku problems with up to 9 missing cells.
The qubit count, $\mathrm{U}3$-gate count, $\mathrm{CX}$-gate count, circuit depth, as well as the runtime on the cloud-based IBM \texttt{simulator\_mps} simulator are shown in Fig.~\ref{fig:benchmarking} and Fig.~\ref{fig:benchmarking_plot}. For 9 empty cells the circuit requires 91 qubits and exhibits a circuit depth of 3968. Note that the qubit count can be reduced further at the expense of the $\mathrm{U}3$-gate count, $\mathrm{CX}$-gate count and circuit depth. This essentially hinges on the implementation of multi-qubit Toffoli gates: The use of ancilla qubits facilitates a reduction in gate count and depth \cite{Amy_2019,Balauca_2022,Seidel_2023}.

\item \textit{Finding a solution.} We verify that a valid solution is found by the \texttt{find\_solution} method for Sudoku instances up to 9 empty fields. As mentioned in Section \ref{sec:backtracking_implementation}, the \texttt{find\_solution} method starts by applying the \texttt{estimate\_phase} function to the entire tree (initialized in the state $\ket{r}$) and, based on the measurement results, recursively applies the function to subtrees in order to find a solution. Specifically, we utilize 10000 shots for quantum phase estimation with precision $2^{-3}$. The root of a new subtree to be explored is then selected from the measured states $\ket{x}\ket{\mathrm{reg}}$ where the phase estimation yields $\ket{\mathrm{reg}}=\ket{0}$, i.e., corresponds to the eigenvalue $1$. This is based on fact that a certain superposition of node states corresponding to a path from the root to a solution is indeed an eigenvector of the operator $R_BR_A$ with eigenvalue $1$ \cite{Montanaro_2016}. 
With this approach, we demonstrate solving Sudoku instances with quantum backtracking. However, 
in order to achieve an advantage in practial applications, parameters, i.e., the precision of the phase estimation and the number of measurements, have to be specified thoroughly \cite{Ambainis_2017}. 

\end{enumerate}

\PreviewEnvironment{tikzpicture}
\begin{figure}[h]
\begin{subfigure}[]{0.5\textwidth}
    \centering
    \begin{tikzpicture}[scale=1.0]

    \begin{scope}
    \draw (0, 0) grid (4, 4);
    \draw[very thick, scale=2] (0, 0) grid (2, 2);

    \setcounter{row}{1}
    \setrow {1}{\color{blue}{2}}{3}{\color{blue}{4}}
    \setrow {3}{\color{blue}{4}}{1}{\color{blue}{2}}

    \setrow {\color{blue}{2}}{1}{\color{blue}{4}}{3}
    \setrow {4}{\color{blue}{3}}{\color{blue}{2}}{\color{blue}{1}}

    \end{scope}

    \end{tikzpicture}
    \subcaption{Solved Sudoku.} 
    \label{fig:sudoku_experiments_unsolved}
\end{subfigure}
\hspace{1cm}

\begin{subfigure}[]{0.5\textwidth}
    \centering
    \begin{tikzpicture}[scale=1.0]

    \begin{scope}
    \draw (0, 0) grid (4, 4);
    \draw[very thick, scale=2] (0, 0) grid (2, 2);

    \setcounter{row}{1}
    \setrow {1}{\color{blue}{$a$}}{3}{\color{blue}{$b$}}
    \setrow {3}{\color{blue}{$c$}}{1}{\color{blue}{$d$}}

    \setrow {\color{blue}{$e$}}{1}{\color{blue}{$f$}}{3}
    \setrow {4}{\color{blue}{$g$}}{\color{blue}{$h$}}{\color{blue}{$i$}}

    \end{scope}

    \end{tikzpicture}
    \subcaption{Unsolved Sudoku problem.} 
    \label{fig:sudoku_experiments_solved}
\end{subfigure}
\caption{Unsolved Sudoku problem and its solution.
}
\label{fig:sudoku_experiments}
\end{figure}

\begin{figure*}[]
\centering
\begin{tabular}{|c|c|c|c|c|c|} 
 \hline
Missing cells & Qubit count & $\mathrm{U}3$-gate count & $\mathrm{CX}$-gate count & Circuit depth & IBM MPS runtime in s\\
\hline
\hline
1 ($a$) & 15 & 1434 & 1157 & 1396 & 5.51\\
\hline
2 ($a,b$) & 22 & 2612 & 2123 & 1732 & 5.33\\
\hline
3 ($a$--$c$) & 29 & 3703 & 2977 & 1979 & 8.17\\
\hline
4 ($a$--$d$) & 40 & 4957 & 3999 & 2127 & 12.95\\
\hline
5 ($a$--$e$) & 46 & 5763 & 4629 & 2266 & 17.00 \\
\hline
6 ($a$--$f$) & 54 & 6944 & 5609 & 2432 & 22.77 \\
\hline
7 ($a$--$g$) & 66 & 8955 & 7303 & 2980 & 40.03 \\
\hline
8 ($a$--$h$) & 75 & 10355 & 8521 & 3270 & 59.93 \\
\hline
9 ($a$--$i$) & 91 & 13074 & 10901 & 3968 & 97.58 \\
\hline
\end{tabular}
\caption{Number of qubits, basis gates and depth for the circuits implementing quantum phase estimation with precision $2^{-3}$ applied to the operator $R_BR_A$, for Sudoku problems with up to 9 missing cells.
The metrics are calculated after transpiling the circuits into the gate set $\{\mathrm{U}3,\mathrm{CX}\}$. Additionally, the runtime of the cloud-based IBM matrix product state simulator \texttt{simulator\_mps} is benchmarked.}
\label{fig:benchmarking}
\end{figure*}

%% file: sections/Conclusion.tex
\section{Conclusion}
\label{sec:conclusion}

In this paper, we provide a short introduction to Montanaro's quantum backtracking algorithm followed by its systematic implementation. To achieve this, we utilize the high-level quantum programming framework \textit{Qrisp}.
The second part of the paper deals with the implementation of the corresponding \texttt{reject} and \texttt{accept} functions, which are necessary for solving Sudoku instances. Even though almost trivial in classical programming, the quantum versions of these functions expose a rather broad spectrum of difficulties. We attribute these difficulties to the following factors:
\begin{enumerate}
    \item \textbf{Limited quantum resources}: Quantum implementations demand utmost efficiency due to the limited computational resources of both simulators and physical quantum backends. To ensure effective software development, the code has to be extremely efficient or otherwise no testing, bug-fixing or bottleneck identification is possible due to a lack of access to verification using simulators.
    \item \textbf{Coherence}: While classical programmers can terminate certain input constellations rather quickly (using for instance \texttt{return} statements), quantum functions are evaluated in superposition. Quantum programmers therefore need to consider every possible input constellation at every single step of the function.
    \item \textbf{Reversibility}: Since quantum computations are reversible by nature, the repertoire of efficient operations compared to irreversible computers is significantly reduced. Additionally, uncomputation (especially if done manually) can pose a considerable challenge for both the software engineer and the quantum resource budget.
    \item \textbf{Lack of tools}: Until very recently, the only way to execute code on a physical backend was to manually plug circuits into each other. This required software engineers to become knowledgeable in many different areas of quantum compilation, yielding a rather limited population of active programmers. This stands in sharp contrast to fields like machine learning, which are also faced with considerable complexities, but have been made accessible to a wide range of users by exposing the right tools.
\end{enumerate}
\textit{Qrisp} as a high-level quantum programming framework directly addresses the last two points. The use of automatic uncomputation \textbf{significantly} simplifies the development and maintenance of multi-level algorithms such as the one presented. Given the complexity in both, the backtracking algorithm and the oracles, a structured approach to quantum programming is inevitable. In this regard, \textit{Qrisp} stands out by enabling systematic development of separate modules, facilitated by its resource management system. In circuit based languages such as \cite{qiskit} or \cite{cirq} the users would have to take care of each and every intermediate ancilla qubit themselves, implying that a scalable implementation would be close to impossible, even if only limited to a Sudoku solver. This observation is confirmed by the glaring lack of compilable code and practical literature on this topic even though the algorithm itself has been known for almost a decade.\\
Meanwhile, the \textit{Qrisp} implementation, with its \texttt{accept}/\texttt{reject} interface, is far from being limited. It is well equipped to tackle a variety of CSPs, further proving the pivotal role that systematic software engineering will play in the future of quantum information science.

%% file: sections/Acknowledgment.tex
This work was funded by the Federal Ministry for Economic Affairs and Climate Action (German: Bundesministerium für Wirtschaft und Klimaschutz) under the project funding number 01MQ22007A and by the European Union under project grant OASEES (HORIZON-CL4-2022, grant agreement no 101092702). The authors are responsible for the content of this publication.

%% file: sections/appendix.tex
\newpage
\section{Implementation of controlled algorithmic primitives}
\label{sec:appendix}
\subsection{Custom controlled routines}
\label{sec:custom_control}
The usage of controlling\footnote{Controlling here means adding a control knob to the circuit, such that the circuit only executes if the qubit of the control knob is in the $\ket{1}$ state.} certain steps of the code in this paper is ubiquitous. A general method for controlling arbitrary quantum circuits is given in \cite{Shende_2006}. As this is a generic one-fits-all method, it is rather inefficient compared to more specific strategies. \\
\textit{Qrisp} allows the programmer to conveniently deploy specific control routines using the \texttt{custom\_control} decorator. This decorator needs to be applied to functions supporting the \texttt{ctrl} keyword. If called within an \texttt{ControlEnvironment}, this keyword argument will be set to the corresponding control qubit of the environment. As an example, we demonstrate how a controlled swap operation can be implemented.

\begin{lstlisting}[language = Python, numbers = none]
from qrisp import *

def regular_swap(a, b):
    cx(a, b)
    cx(b, a)
    cx(a, b)

@custom_control
def custom_controlled_swap(a, b, ctrl = None):

    # Create QuantumEnvironment to wrap the 
    # middle CX
    if ctrl is None:
        # If no control value is given, the 
        # function is not called in controlled 
        # mode, therefore  the environment 
        # doesn't need to do anything.
        env = QuantumEnvironment()
    else:
        # In the case that a control qubit is 
        # given, the environment will be the 
        # ControlEnvironment to compile the 
        # generic control routine of CX.
        env = control(ctrl)

    cx(a, b)
    # Call the middle CX in the environment
    with env:
        cx(b, a)
    cx(a, b)

###########
# Testing #
###########

a = QuantumBool()
b = QuantumBool()
c = QuantumBool()

# Call the previously created functions inside a
# ControlEnvironment
with control(c):
    regular_swap(a, b)
    barrier([a,b,c])
    custom_controlled_swap(a, b)

\end{lstlisting}

This code first defines a function that performs a regular two qubit swapping operation by sequentially applying three CX gates. Using the generic control routine, it is possible to turn this function into its controlled version (the so called Fredkin-gate) by controlling each CX gate. A much more efficient version is, however, possible by leveraging the fact that only the middle CX needs to be controlled. This is because if the control qubit is in the $\ket{0}$ state, the middle CX will not be executed, so the outer CX's cancel each other resulting in a net-zero operation. We achieve this behavior by setting up a \texttt{QuantumEnvironment}\footnote{\texttt{QuantumEnvironment}s are a central language construct specified by \textit{Qrisp}. They allow the user to apply certain modes of compilation to the code within the environment. Important examples of such environments are the \texttt{ControlEnvironments}, which control the content, the \texttt{InversionEnvironment}, which inverts/daggers the content and the baseclass, which does simply nothing}, which wraps the middle CX in both, controlled and not controlled cases.

\begin{itemize}
    \item In case the swap function was called with \texttt{ctrl = None}, we know that the regular version of this function is required. Therefore, we use the base \texttt{QuantumEnvironment}, which performs no non-trival compilation steps and simply returns its content.
    \item For the controlled case, we use the \texttt{ControlEnvironment}, which calls the generic control routine, implying a Toffoli gate is compiled.
\end{itemize}
If instead of the CX, another function was called inside the environment and that function is exposing a \texttt{custom\_control} itself, then the controlled version of this function would be called. In other words: Nested \texttt{custom\_control} structures are possible.
Compiling this script yields the circuit in Fig.~\ref{fig:custom_control circuit}.
\begin{figure}
\scalebox{1.0}{
\Qcircuit @C=1.0em @R=0.8em @!R { \\
	 	\nghost{{a.0} :  } & \lstick{{a.0} :  } & \ctrl{1} & \targ & \ctrl{1} \barrier[0em]{2} & \qw & \ctrl{1} & \targ & \ctrl{1} & \qw & \qw\\
	 	\nghost{{b.0} :  } & \lstick{{b.0} :  } & \targ & \ctrl{-1} & \targ & \qw & \targ & \ctrl{-1} & \targ & \qw & \qw\\
	 	\nghost{{c.0} :  } & \lstick{{c.0} :  } & \ctrl{-1} & \ctrl{-1} & \ctrl{-1} & \qw & \qw & \ctrl{-1} & \qw & \qw & \qw\\
\\ }}
\caption{The compiled circuit of the script in section \ref{sec:custom_control}. We see that the custom controlled function performs the same unitary as the generic one but with less resources required. Regarding the CX count we have: 18 compared to 8.}\label{fig:custom_control circuit}
\end{figure}

\subsection{Custom controlled \texttt{qstep\_diffuser}}
\label{appendix:qstep}
Since the quantum step operator is subject to quantum phase estimation, we also need the controlled version of it. The generic control routine would result in quite a lot of overhead (especially compared to Grover iterations). Fortunately, our algorithm can be formulated in the form of a \textit{conjugation}. That is, the unitary takes the form:
\begin{align}
    W_{\text{step}} = U V U^{\dagger}
\end{align}
where $U$ is the unitary of the \texttt{psi\_prep} function and $V$ is the unitary of the MCZ gates described in algorithm \ref{alg:qstep_diffuser}.

Circuit structure like these can be efficiently controlled using:
\begin{align}
    cW = U cV U^{\dagger}
\end{align}
In other words: Only the center operation needs to be controlled because if it is not executing, the other two operators cancel each other out. We can therefore effectively control the quantum step operator by only adding extra controls to the MCZ gates. This implies barely any overhead as \textit{Qrisp} automatically utilizes the ancilla qubits that have been recycled from the oracle calls to synthesize highly efficient MCZ implementations, scaling only logarithmically in depth \cite{Balauca_2022}.

\subsection{Custom controlled XX+YY}
\label{sec:custom_control_xxyy}
The XX+YY gate with $\beta = \frac{\pi}{2}$ can be decomposed to CNOT and single qubit rotations in the following way \cite{Anselmetti_2021}:
\begin{center}
\scalebox{1.0}{
\Qcircuit @C=1.0em @R=0.2em @!R { \\
	 	\nghost{{qb.0} :  } & \lstick{{qb.0} :  } & \qw & \targ & \gate{\mathrm{R_Y}\,(-\phi/2)} & \targ & \qw & \qw & \qw\\
	 	\nghost{{qb.1} :  } & \lstick{{qb.1} :  } & \gate{\mathrm{H}} & \ctrl{-1} & \gate{\mathrm{R_Y}\,(-\phi/2)} & \ctrl{-1} & \gate{\mathrm{H}} & \qw & \qw\\
\\ }}
\end{center}
Since this is also a conjugation, a more efficient procedure for controlling is obvious: Simply control the RY gates. There are also more improvements achievable.

The following circuit was numerically verified to yield the equivalent output states for all inputs but the $\ket{11}$. This input is essentially irrelevant because the \texttt{h} variable is one hot encoded and can never contain two 1s.
\begin{center}
\hspace*{-0.5cm}
\scalebox{0.9}{
\Qcircuit @C=1.0em @R=0.2em @!R { \\
	 	\nghost{{qb.0} :  } & \lstick{{qb.0} :  } & \qw & \targ & \gate{\mathrm{R_Y}\,(\gamma)} & \gate{\mathrm{H}} & \gate{\mathrm{R_Y}\,(-\gamma)} & \targ & \qw & \qw \\
	 	\nghost{{qb.1} :  } & \lstick{{qb.1} :  } & \gate{\mathrm{H}} & \ctrl{-1} & \gate{\mathrm{R_Y}\,(\gamma)} & \gate{\mathrm{H}} & \gate{\mathrm{R_Y}\,(-\gamma)} & \ctrl{-1} & \gate{\mathrm{H}} & \qw \
\\ }}
\end{center}
Here,
\begin{align}
    \gamma = \frac{\phi+\pi}{4}.    
\end{align}
The advantage of this form is that we can control an H gate instead of an RY gate. Controlled H gates can be realized with the following circuit and therefore require only a single CNOT \cite{Amy_2013}, whereas controlled RY gates require double the amount:
\begin{center}
\Qcircuit @C=1.0em @R=0.2em @!R { \\
	 	\nghost{{qb.0} :  } & \lstick{{ctrl} :  } & \qw & \qw & \qw & \ctrl{1} & \qw & \qw & \qw & \qw & \qw\\
	 	\nghost{{qb.1} :  } & \lstick{{trgt} :  } & \gate{\mathrm{S}} & \gate{\mathrm{H}} & \gate{\mathrm{T}} & \targ & \gate{\mathrm{T^\dagger}} & \gate{\mathrm{H}} & \gate{\mathrm{S^\dagger}} & \qw & \qw\\
\\ }
\end{center}

Another additional improvement is possible in situations when we are not treating a binary backtracking tree, that is, if each node has more than two children. In this case, the variables in \texttt{branch\_qa} have multiple qubits and therefore the XX+YY gate requires multiple controls. Instead of having two $n$-controlled X gates, we can use the following circuit:
\begin{center}
\hspace*{-0.5cm}
\scalebox{0.9}{
\Qcircuit @C=1.0em @R=0.2em @!R { \\
	 	\nghost{{ctrl.0} :  } & \lstick{{ctrl.0} :  } & \qw & \qw & \qw & \qw & \ctrl{1} & \qw & \qw & \qw & \qw & \qw & \qw\\
	 	\nghost{{ctrl.1} :  } & \lstick{{ctrl.1} :  } & \qw & \qw & \qw & \qw & \ctrl{1} & \qw & \qw & \qw & \qw & \qw & \qw\\
	 	\nghost{{qb.0} :  } & \lstick{{qb.0} :  } & \gate{\mathrm{S}} & \gate{\mathrm{H}} & \gate{\mathrm{T}} & \ctrl{1} & \targ & \ctrl{1} & \gate{\mathrm{T^\dagger}} & \gate{\mathrm{H}} & \gate{\mathrm{S^\dagger}} & \qw & \qw\\
	 	\nghost{{qb.1} :  } & \lstick{{qb.1} :  } & \gate{\mathrm{S}} & \gate{\mathrm{H}} & \gate{\mathrm{T}} & \targ & \qw & \targ & \gate{\mathrm{T^\dagger}} & \gate{\mathrm{H}} & \gate{\mathrm{S^\dagger}} & \qw & \qw\\
\\ }}
\end{center}
\subsection{Custom controlled \textit{classical-quantum} equality check}
\label{sec:controled_cq_comparison}
As elaborated upon earlier, the \textit{classical-quantum} comparison is performed by batching together multiple classical values belonging to a single quantum value, and evaluating the batches of comparisons using a \texttt{QuantumDictionary}.
In \textit{Qrisp}, this data structure utilizes quantum logic synthesis - to be specific, the algorithm presented in \cite{Seidel_2023}. The CNOT count and quantum circuit depth are proportional to $\approx m 2^n$ where $n$ is the bit-width of the keys and m is the bit-width of the values. Applying the generic control procedure \cite{Shende_2006} would require all CNOT gates to be turned into Toffolis, implying an increase in CNOT count by a factor of $\times 6$ (using the Toffoli implementation from \cite{Amy_2013}). A much more efficient method for performing controlled quantum logic synthesis is, however, to call the synthesis algorithm again, but with the extra input bit \texttt{ctrl}, such that every output with \texttt{ctrl} $= 0$ is also $0$. According to the above formula, the overhead in CNOT count is therefore only $\times 2$. Note that this technique is not only limited to Gray-code traversal based quantum logic synthesis, but \textit{any} logic-synthesis algorithm. Turn to Fig.~\ref{fig:lsynthtable} for an example.

\begin{figure}[h!]
    \begin{subfigure}[c]{0.22\textwidth}
    \scalebox{0.8}{
    \begin{tabular}{|c|c||c|}
        \hline
        \textbf{$x_0$} & \textbf{$x_1$} & \textbf{$f(x)$} \\
        \hline
        0 & 0 & 1 \\
        0 & 1 & 0 \\
        1 & 0 & 1 \\
        1 & 1 & 1 \\
        \hline
    \end{tabular}}
    \subcaption{}
    \label{tbl:lsyntha}
    \end{subfigure}
    \begin{subfigure}[c]{0.22\textwidth}
    \scalebox{0.8}{
    \begin{tabular}{|c|c|c||c|}
        \hline
        \textbf{$ctrl$}& \textbf{$x_0$} & \textbf{$x_1$} & \textbf{$cf(x)$} \\
        \hline
        0 & 0 & 0 & 0 \\
        0 & 0 & 1 & 0 \\
        0 & 1 & 0 & 0 \\
        0 & 1 & 1 & 0 \\
        1 & 0 & 0 & 1 \\
        1 & 0 & 1 & 0 \\
        1 & 1 & 0 & 1 \\
        1 & 1 & 1 & 1 \\
        \hline
    \end{tabular}}
    \subcaption{}
    \label{tbl:lsynthb}
    \end{subfigure}
    \caption{Table \ref{tbl:lsyntha}: A random truth table, which could be synthesized with quantum logic synthesis. Table \ref{tbl:lsynthb}: A truth table representing the controlled call of the previous table.}
    \label{fig:lsynthtable}
\end{figure}

\subsection{Custom controlled quantum-quantum equality check}
\label{sec:controled_qq_comparison}
To compare two \texttt{QuantumVariable}s \texttt{a} and \texttt{b} in \textit{Qrisp}, the following procedure is followed:

\begin{enumerate}
    \item CNOT from each qubit of \texttt{a} into the corresponding qubit of \texttt{b}.
    \item MCX with control state 0, controlled on \texttt{b} onto the result qubit.
    \item Repeat step 1 to restore the original value of \texttt{b}.
\end{enumerate}
The corresponding circuit looks like this:
\begin{center}
\scalebox{1.0}{
\Qcircuit @C=1.0em @R=0.2em @!R { \\
	 	\nghost{{a.0} :  } & \lstick{{a.0} :  } & \ctrl{3} & \qw & \qw & \qw & \qw & \qw & \ctrl{3} & \qw & \qw\\
	 	\nghost{{a.1} :  } & \lstick{{a.1} :  } & \qw & \ctrl{3} & \qw & \qw & \qw & \ctrl{3} & \qw & \qw & \qw\\
	 	\nghost{{a.2} :  } & \lstick{{a.2} :  } & \qw & \qw & \ctrl{3} & \qw & \ctrl{3} & \qw & \qw & \qw & \qw\\
	 	\nghost{{b.0} :  } & \lstick{{b.0} :  } & \targ & \qw & \qw & \ctrlo{1} & \qw & \qw & \targ & \qw & \qw\\
	 	\nghost{{b.1} :  } & \lstick{{b.1} :  } & \qw & \targ & \qw & \ctrlo{1} & \qw & \targ & \qw & \qw & \qw\\
	 	\nghost{{b.2} :  } & \lstick{{b.2} :  } & \qw & \qw & \targ & \ctrlo{1} & \targ & \qw & \qw & \qw & \qw\\
	 	\nghost{{res.0} :  } & \lstick{{res} :  } & \qw & \qw & \qw & \targ & \qw & \qw & \qw & \qw & \qw\\
\\ }}
\end{center}
This circuit performs the desired function (equality check) because the multi-controlled X gate in the middle only triggers if each qubit of \texttt{b} is in the $\ket{0}$ state. This however is only the case if \texttt{a} and \texttt{b} agree on every qubit.

This circuit is a conjugation as described above, therefore the controlled version is:
\begin{center}
\scalebox{1.0}{
\Qcircuit @C=1.0em @R=0.2em @!R { \\
	 	\nghost{{ctrl.0} :  } & \lstick{{ctrl} :  } & \qw & \qw & \qw & \ctrl{4} & \qw & \qw & \qw & \qw & \qw\\
	 	\nghost{{a.0} :  } & \lstick{{a.0} :  } & \ctrl{3} & \qw & \qw & \qw & \qw & \qw & \ctrl{3} & \qw & \qw\\
	 	\nghost{{a.1} :  } & \lstick{{a.1} :  } & \qw & \ctrl{3} & \qw & \qw & \qw & \ctrl{3} & \qw & \qw & \qw\\
	 	\nghost{{a.2} :  } & \lstick{{a.2} :  } & \qw & \qw & \ctrl{3} & \qw & \ctrl{3} & \qw & \qw & \qw & \qw\\
	 	\nghost{{b.0} :  } & \lstick{{b.0} :  } & \targ & \qw & \qw & \ctrlo{1} & \qw & \qw & \targ & \qw & \qw\\
	 	\nghost{{b.1} :  } & \lstick{{b.1} :  } & \qw & \targ & \qw & \ctrlo{1} & \qw & \targ & \qw & \qw & \qw\\
	 	\nghost{{b.2} :  } & \lstick{{b.2} :  } & \qw & \qw & \targ & \ctrlo{1} & \targ & \qw & \qw & \qw & \qw\\
	 	\nghost{{res.0} :  } & \lstick{{res} :  } & \qw & \qw & \qw & \targ & \qw & \qw & \qw & \qw & \qw\\
\\ }}
\end{center}
In practice, the \textit{Qrisp} compiler synthesizes the multi-controlled X gate using several ancillae \cite{Balauca_2022}, which become available from prior deallocations, resulting in barely any overhead for the extra control-knob.

\section{Phase tolerant synthesis}
\label{sec:pt_compilation}

Phase tolerant compilation is a technique that is applicable in situations, where it is known a priori that a certain function will be uncomputed at some point in the future. This is based on the fact that for certain quantum functions of the form

\begin{equation}
    U_f \ket{x}\ket{0} = \ket{x}\ket{f(x)},
\end{equation}
it is sometimes more efficient to compile a circuit with "garbage" phases:

\begin{equation}
    \hat{U}_f \ket{x}\ket{0} = \text{exp}(i \phi_x) \ket{x}\ket{f(x)}
\end{equation}

If we know that such a function, at some point in the future, will be uncomputed by calling the inverse on the same inputs, the garbage phases cancel out, implying the net function has the right phase behavior. Since we can lower our requirements to the compilation of such functions, we call this method of compilation "phase tolerant". For 2 and 3-controlled X gates, phase tolerant versions have been given in \cite{Maslov_2016}. For more controls and especially for logic synthesis, phase tolerant circuit constructions can be found in \cite{Seidel_2023}.

With regard to \textit{Qrisp}, in many situations the automatic uncomputation algorithm \cite{Seidel_2023_uncomputation, Paradis_2021} is able to detect when phase tolerant replacement is possible and automatically generates the corresponding circuit. In some situations, this does not work out so the programmer has to specify this behavior manually. 

For the Sudoku context, both comparison types profit a lot from these features. For the quantum-quantum comparison, we can write a new comparison function:

\begin{lstlisting}[language = Python, numbers = none]
@custom_control
def pt_comparison(a, b, ctrl = None):

    control_qb_list = list(b)
    if ctrl is not None:
        control_qb_list.append(ctrl)
    
    res = QuantumBool()

    cx(a, b)
    mcx(control_qb_list, res, method = "gray_pt")
    cx(a, b)
    
    return res

\end{lstlisting}

The keyword argument \texttt{method} indicates that instead of the default MCX algorithm, the phase tolerant version should be used. This transformation is valid because we know that the comparison value will be uncomputed later and therefore the additional "garbage phases" cancel out. Compared to the naive version where a 3-controlled X gate would be called (requiring 14 CX's), the phase tolerant version given in \cite{Maslov_2016} requires only 6 CX's.

For the classical-quantum comparisons, this technique is also applicable, considering the logic synthesis algorithm is based on \cite{Seidel_2023}. For this, we need even less modification: In the function \texttt{eval\_cq\_checks}, it suffices to simply replace the inner workings of the loop to

\begin{lstlisting}[language = Python, numbers = none]
with control(height[key], ctrl_method = "gray_pt"):
    eq_qbl = cq_eq_check(branch_qa[key], values)
\end{lstlisting}
For a 2 qubit evaluation this reduces the number of required CXs from 16 to 10. Since both comparison circuits are basically an alternating sequence of CX and single qubit rotations, these optimizations equally reduce the single qubit gate count, especially the T-count.